\documentclass[useAMS,usenatbib,twocolumn]{mn2e}

\usepackage{amsmath,amssymb}
\usepackage{subfigure}
\usepackage{graphicx}
\usepackage{natbib}
\usepackage[hyphens]{url}

\newcommand{\spose}[1]{\hbox to 0pt{#1\hss}}
\newcommand{\lta}{\mathrel{\spose{\lower 3pt\hbox{$\mathchar"218$}}
     \raise 2.0pt\hbox{$\mathchar"13C$}}}
\newcommand{\gta}{\mathrel{\spose{\lower 3pt\hbox{$\mathchar"218$}}
     \raise 2.0pt\hbox{$\mathchar"13E$}}}

\newcommand{\pc}{\mbox{\,pc}}

\newcommand{\kpc}{\mbox{\,kpc}}
\newcommand{\kms}{\mbox{\,km/s}}
\newcommand{\myr}{\mbox{\,Myr}}

\newcommand{\cm}{\mbox{\ cm}}
\newcommand{\msun}{\mbox{\,M}_\odot}

\newcommand{\aj}{AJ}         
\newcommand{\apj}{ApJ}       
\newcommand{\apjl}{ApJL}     
\newcommand{\mnras}{MNRAS}   
\newcommand{\aap}{A\&A}      

\title[Ram Pressure Dynamics]{Direct Simulation Monte Carlo for
  astrophysical flows: II. Ram pressure dynamics}
\author[M.~D.~Weinberg]{Martin D. Weinberg\thanks{E-mail:
    weinberg@astro.umass.edu}\footnotemark[1] \\
  Department of Astronomy, University of Massachusetts, Amherst,
  MA 01003-9305, USA }

\pagerange{\pageref{firstpage}--\pageref{lastpage}}
\pubyear{2013}

\begin{document}

\label{firstpage}

\date{\today}
\pagerange{\pageref{firstpage}--\pageref{lastpage}} \pubyear{2013}

\maketitle

\begin{abstract}
  We use the \emph{Direct Simulation Monte Carlo} (DSMC) method
  combined with an n-body code to study the dynamics of the
  interaction between a gas-rich spiral galaxy and intracluster or
  intragroup medium, often known as the ram pressure scenario. The
  advantage of this gas kinetic approach over traditional
  hydrodynamics is explicit treatment of the interface between the hot
  and cold, dense and rarefied media typical of astrophysical flows
  and the explicit conservation of energy and momentum and the
  interface.  This approach yields some new physical insight.  Owing
  to the shock and backward wave that forms at the point ICM--ISM
  contact, ICM gas is compressed, heated and slowed.  The shock
  morphology is Mach-disk-like. In the outer galaxy, the hot turbulent
  post-shock gas flows around the galaxy disk, while heating and
  ablating the initially cool disk gas.  The outer gas and angular
  momentum are lost to the flow.  In the inner galaxy, the hot gas
  pressurises the neutral ISM gas causing a strong two-phase
  instability.  As a result, the momentum of the wind is no longer
  impulsively communicated to the cold gas as assumed in the Gunn-Gott
  (1972) formula, but oozes through the porous disk, transferring its
  linear momentum to the disk en masse.  The escaping gas mixture has
  a net positive angular momentum and forms a slowly rotating sheath.
  The shear flow caused by the post-shock ICM flowing through the
  porous multiphase ISM creates a strong Kelvin-Helmholtz instability
  in the disk that results in Cartwheel-like ring and spoke
  morphology.
\end{abstract}


\section{Introduction}
\label{sec:intro}

The interaction between the intracluster medium (ICM) or the
intragroup medium (IGM) and the interstellar medium (ISM) in a
gas-rich spiral is often identified as the source of star bursts and
gas starvation followed by morphological transformation.  Recent
observations of this process in the Virgo cluster have spawned a
number of detailed hydrodynamic simulations that investigate the
conditions required for these scenarios.  This paper applies the
hybrid Direct Simulation Monte Carlo (DSMC) method from
\citet[hereafter Paper 1]{Weinberg:13a} to the collision of two
distinct gas components with different densities, temperatures and
bulk velocities.  Such an interaction is typically referred to as
\emph{ram pressure}.  Many if not most ram pressure interactions are
in the \emph{transitional} regime which is not quite hydrodynamic.
The subsequent evolution of the system depends on the dynamics at the
collision interface and thereby provides a good test of the kinetic
approach.  Thus, the advantages of DSMC all come to the fore when
considering the ram-pressure mechanism and make ram pressure both a
good problem for code comparison.

Although DSMC was developed for rarefied and transitional gas flows,
the method has advantages for near continuum (i.e. low Knudsen number)
flows: 1) there is no Courant-Friedrichs-Lewy (CFL) condition for
DSMC; DSMC is unconditionally stable; 2) the method naturally
resolves shocks without need for artificial viscosity or shock-capturing 
techniques; specifically, the inter-particle interactions on the
mean-free-path scale are explicitly resolved and, therefore, there are
no formal contact discontinuities; and 3) the DSMC approach naturally
resolves the gas-phase interfaces that plague many hydrodynamic
simulations when applied to scenarios with ISM cooling, e.g. giving
rise to the \emph{over-cooling} problem.  The penalty for these
advantages is much lower computational efficiency than the same
computation using traditional grid hydrodynamics.  This approach
requires many more gas particles in the continuum limit than mesh
cells or SPH particles, although it has the advantage of consistently
transitioning from dense to dilute gases, fully capturing shocks, and
resolving phase boundaries. By design, DSMC is correct even as the
mean free path becomes large.  Such regimes are common in intracluster
gas dynamics, hot bubbles and conduction at interfaces, and in
multiphase ISM on small scales which require accurate treatment of
interfaces.  Moreover, this method maintains the underlying physics of
the multiphase interfaces.

As the work proceeded, it became clear that the standard explanation
of ram pressure did not explain results of the numerical experiments,
either those described here or published elsewhere.  To illustrate
these differences, we will begin with the standard explanation for
astrophysical ram pressure in Section \ref{sec:ram_pres}, followed by an
exploratory sequence of simulations of increasing generality in
Section \ref{sec:experiments}.  These results will be compared with
published observational and theoretical findings in
Section \ref{sec:discuss}.  In brief, many of the features found here have
been identified and reported by others (as expected for a code test)
but the simulations here reveal a complexity of multiphase
instabilities and angular momentum transport in the cold gas disk that
has not been described previously.  The overall explanation most
commonly offered for this interaction is the ram-pressure paradigm
envisioned by \citet[hereafter Gunn-Gott]{Gunn.Gott:1972}.  However,
we will argue that the dominant processes driving the evolution of a
galaxy with a wind is not Gunn-Gott ram pressure but the combination
of shock heating, boundary-layer ablation, Kelvin-Helmholtz
instability in a multiphase medium, and angular momentum transfer.
Because of this complexity, any specific observed interaction is
likely to depend on the details of the galaxy, its gravitational
environment and the density and speed of the wind.  Owing to the
expense of the DSMC method, we can not afford an exhaustive survey,
but choose a single galaxy model and a high-density wind of modest to
low velocity.  In particular, we explore and describe the results for
a one-dimensional slab, a three-dimensional disk + halo model with
face-on, oblique, and edge-on winds, and include a case with and
without a hot, coronal halo.  We summarise in Section \ref{sec:summary}.

\section{Ram pressure}
\label{sec:ram_pres}

\subsection{The astronomical setting}
\label{sec:ram_astro}

Galaxy clusters and groups are often permeated by hot, tenuous, X-ray
emitting gas known as the intracluster medium (ICM). Throughout this
paper, we will refer to both the intracluster medium and the
intragroup medium (IGM) as the ICM.  The pressure exerted by flows in
this medium on a cold gas disk bound to the stellar disk is often
described as ram pressure.  In physics and engineering, ram pressure
denotes the back pressure applied to a solid body moving through a
fluid: $P=\rho v$ where $P$ is the pressure, $\rho$ is the mass
density, and $v$ is the relative speed of the body.  Astronomers use
\emph{ram pressure} to denote the pressure applied by one fluid on
another when the two fluids have distinct phases and significant
relative bulk velocity.  Let us refer to a hot tenuous gas that has
significant bulk velocity with respect to the galaxy disk as a
\emph{wind}.  \emph{Ram pressure stripping} is said to occur if this
wind is able to impart sufficient momentum to the cold disk that it
overcome the gravitational attraction of the galaxy.  Although we will
retain the traditional terminology, this paper will demonstrate that
\emph{ram pressure} is a physical misnomer in the astrophysical
context; momentum transfer \emph{does} occur, but it does not act like
a macroscopic blunt body moving in a hydrodynamic flow.  Rather, the
details of the interaction of the two gas components with differing
densities and temperatures is sufficiently dynamic that the ram
pressure paradigm misses many of the essential features.

Regardless of the dynamical details, observational evidence for the
interaction clearly exists.  For simplicity, let us follow the
standard practice of referring to the flow in hot medium relative to
the frame of the cold galaxy as a \emph{wind} and the resulting
momentum transfer as \emph{ram pressure}.  Evidence for ram pressure
can be found in many galaxy clusters. For example, three morphological
features in NGC 4402\footnote{Hubble image:
  \url{http://www.spacetelescope.org/images/heic0911c}} suggest strong
gas interactions.  First, the bowed morphology of the gas suggests a
balance between the ram pressure and the lower gravitational force at
larger disk radii.  Secondly, the lower gas content in the extended
blue stellar disk suggests a recent epoch of star formation,
presumably induced by the interaction event.  Finally, lanes of
reddening can be seen trailing the inferred motion of galaxy in the
cluster.  The expected result of ram pressure stripping is a galaxy
which contains very little cold gas. This effectively halts star
formation in the galaxy, supporting the belief that ram pressure
stripping could be one of the processes responsible for the
morphology--density relation \citep[see][and references
therein]{VanderWel.etal:2010}.  \citet{Crowl.etal:2007} describe the
observational case for ram-pressure stripping of NGC 4402 in detail.

Another interesting candidate is NGC 4522\footnote{ Hubble image:
  \url{http://www.spacetelescope.org/images/heic0911b}}.
Line-of-sight velocities suggest that this galaxy is currently falling
towards the centre of the Virgo cluster (toward the bottom of the HST
image\footnotemark[2]). The wind direction is less
certain but presumably from below.  The bowed and truncated disk, and
the concentration of dust and gas to one side of the galaxy are all
indicators that ram pressure stripping is forcing gas out of the
galaxy.  H-alpha images of this galaxy suggest that there is no gas in
the outer disk, but there are gas filaments emerging from the plane of
the disk, presumably caused by the interaction with the intracluster
medium.  The plume geometry of these features in NGC 4522 is
consistent with predictions by the simulations described in
Section \ref{sec:gal_hard} that post-shock gas will excite and then ooze
through multiphase instabilities in the cold disk.

\subsection{The Gunn--Gott ram-pressure criterion}
\label{sec:gunn_gott}

We review the physical arguments for the classic Gunn-Gott ram
pressure formula before moving on to the numerical experiments using
DSMC in the next section.  Consider a parcel of wind gas moving with
respect the cold disk. The momentum per unit volume in the wind is
\begin{equation}
  \mbox{wind momentum} = \rho_{ICM} v_{wind}.
\end{equation}
The wind parcel crosses the stellar and gaseous disk in the time
interval approximately given by $h/v_{wind}$.  Therefore, the momentum
delivered by the stream of ICM parcels to the cold gas disk is
\begin{equation}
  \mbox{wind force} = \rho_{ICM} v_{wind}
  \times\underbrace{v_{wind}/h}_{\mbox{\tiny
      1/duration}}.
\end{equation}
The horizontally bracketed term in the equation above is the arrival
frequency of ICM gas parcels, ignoring any dissipation.  Finally, the
condition that the volume force applied by the wind exceeds the
gravitational restoring force binding the cold gas to the stellar disk
is
\begin{equation}
  \rho_{ICM} v_{wind}^2/h >
    \rho_{gas}\frac{\partial\Phi}{\partial z}.
\end{equation}
Rearranging, we get a condition on the ram pressure required to strip
the disk gas:
\begin{equation}  
  P_{ram} = \rho_{ICM} v_{wind}^2 >
  \Sigma_{gas}\frac{\partial\Phi}{\partial z},
  \label{eq:ggc}
\end{equation}
where $\Sigma_{gas}=\int dz\,\rho_{gas}(z)\approx \rho_{gas}(0)h$ is
the gas surface density.

\section{Experiments}
\label{sec:experiments}

We will attempt to verify this physical picture with a short series of
numerical experiments of increasing complexity.  The numerical methods
described in Paper 1 are briefly reviewed in the first section below
and the initial conditions for the experiments are presented in
Section \ref{sec:ics}.  These are followed by in-depth descriptions of the
one-dimensional (Section \ref{sec:simple_slab}) and three-dimensional
(Section \ref{sec:gal_hard}) hot wind--cold disk simulations.

\subsection{Methodology}
\label{sec:DSMC}

DSMC solves the collisional Boltzmann equation when the collision term
is dominated by two-body collisions for one or more species with or
without internal degrees of freedom.  Unlike molecular dynamics, the
solution is obtained by sequentially computing the collisionless or
\emph{ballistic} flow followed by simulating the collisions implied by
the Boltzmann collisional term.  Essentially, the latter step is a
Monte Carlo evaluation of the collision integral, and therefore a
single simulation is one realisation of an ensemble of possible
solutions.  To help prevent the near-continuum limit from being a
computational bottleneck, Paper 1 describes a hybrid DSMC approach
that recovers the Navier-Stokes equation limit for mean free paths
that are very small compared to any flow scale.  This is achieved by
limiting the number of collisions when the mean-free path becomes very
small compared to any other scale of interest.  In this limit, the
solution in each collision cell is replaced by an equivalent
thermalised randomly-generated distribution that conserves energy and
momentum; this is called the Equilibrium Particle Simulation Method
\citep[EPSM,][]{Pullin:80,Macrossan:2001}.  The hybrid DSMC-EPSM
algorithm was parallelised and merged with a collisionless N-body
solver called EXP.  Paper 1 demonstrates that this code correctly
reproduces the standard shock-tube tests and develops Kelvin-Helmholtz
instabilities following the analytic dispersion relation.

In the simulations below, the stellar disk and dark matter halo are
self-gravitating and mutually interacting.  The gas feels the
gravitational field generated by the stellar and dark matter
components but do not exert gravitational force on these components or
on itself.  Therefore, in these runs, the gas cannot be
gravitationally unstable.  This is \emph{not} a fundamental limitation
of the combined EXP-DSMC code.  For example, EXP also includes a
PIC-Fourier Poisson solver that could be used to investigate the
small-scale gravitational response of the gas to the momentum transfer
from the wind.  Rather, because the gravitational instabilities in the
gas are likely to be on a much smaller scale than those of the stellar
and dark matter components and because the mass in diffuse gas is very
small compared to the mass in the stellar disk or halo, we choose to
focus on the large-scale features of the flow.  A simulation with a
coronal halo is briefly described in Section \ref{sec:hothalo}; the heat
generated by the wind interaction is sufficient to prevent the coronal
gas from cooling.  All of the DSMC simulations described here are
fully three-dimensional, even though the initial conditions and
gravitational Poisson solver have explicit symmetries.

\subsection{Initial conditions}
\label{sec:ics}

The first set of experiments (described in detail in
Section \ref{sec:simple_slab}) explores a one-dimensional wind interaction
with a slab of cold gas confined by stars and dark matter.  A planar
stellar slab is realised from the exact isothermal solution to the
equations of hydrostatic equilibrium.  In the second set of
experiments (Section \ref{sec:gal_hard}), the stellar disk and halo are
initialised as described in \citet{Holley-Bockelmann.Weinberg.ea:05}.
In brief, the steps in realising the phase-space distribution for the
second set of simulations are as follows: 1) the dark halo
distribution function is realised by the Eddington inversion method
\citep[e.g.][]{Binney.Tremaine:87}, modified to include the monopole
component of the stellar disk.  The stellar disk is realised using
Jeans' equations with the potential of the disk and halo particles as
computed by the EXP Poisson solver.  The profiles are Milky-Way like:
the dark halo is an NFW \citep{Navarro.Frenk.ea:96} profile with
$c=15$ and $R_{vir}=300\kpc$ and an exponential stellar disk with
scale length of $3\kpc$ and scale height of $350\pc$.  The
stellar-mass to dark-matter ratio is 1/15.  In both cases, the cold
gas disk is initially in local hydrostatic equilibrium with a
temperature of 3000 K based on numerical solutions of the
one-dimensional Euler equations.  For the Milky-Way like model, the
gas disk has the same scale length as the stars and is rotationally
supported.  We do not include a separate bulge or spheroidal component
in this model.

As described in Paper 1, these tests use an LTE cooling approximation
rather than the full cross-section based Monte Carlo simulation of the
Boltzmann collision term to facilitate a comparison with most of the
published work on the subject.  This model \citep{Black:81} includes
atomic processes only so that temperatures lower than 3000 K will be
of little energetic consequence.  We will describe gas whose
temperature is at the minimum of the cooling curve `cold' gas.  Use of
the full DSMC method allows gas of any chemical composition and
excitation state, and this will be studied in a later contribution
(see Section \ref{sec:summary} for additional details).  The hot halo
used in Section \ref{sec:hothalo} is realised for simulation solving
the hydrostatic equilibrium equation for the monopole term of the
total gravitational potential fixing the temperature to $T=10^6$ K and
the density at the solar circle to $n=10^{-4}$ atom/cc.  The wind is
realised as a uniform column with $T=10^6$ K and $n=10^{-3} $ atom/cc,
initially moving toward the galaxy with a uniform bulk velocity of
$2\times V_{vir}=240\kms$.  This modest wind velocity is low compared
to that expected in a large cluster investigated by some authors (see
Section \ref{sec:discuss}).  However, the momentum and energy of the
wind is carried by particles in a DSMC simulation.  Therefore,
simulating a high-velocity wind requires a large column particles.
This is computationally expensive, so we compromise with a wind speed
more typical of a group interaction to lengthen the duration of the
interaction. We explore three wind directions relative to the rotation
axis of the disk: 0, 45 and 90 degrees.  Once more than approximately
75\% of the column has passed the galaxy, the simulation is
terminated.  The duration of the simulation, then, is approximately
400 Myr in our physical units.

The code is heavily instrumented, accumulating the mean Knudsen
number, comparing the mean ballistic trajectory to the cell size,
collisions per simulation particle, fraction of particles and cells in
the EPSM limit for every cell at every time step.  The mean velocity
field, $\langle\mathbf{v}\rangle$, is characterised by gridding the
gas particles and computing streamlines by numerically solving the
first-order differential equations for the flow: $\mathbf{\dot
  x}=\langle\mathbf{v}\rangle$. Other mean quantities of the flow such
as energy and angular momentum fluxes are readily computed from the
gas particle distribution.  These ensemble diagnostics will be used to
illustrate the field quantities in the following sections.  However,
unlike hydrodynamic simulations, the particles in a DSMC simulation
are \emph{not} tracers of the underlying fields but carry the
intrinsic quantities of the microscopic physical state of the gas.
The computation of ensemble quantities, then, artificially degrades
the true resolution of the simulation.  In fact, the gas dynamics in
the simulation takes place on a scale much smaller than the ensemble
grid scale, often by an order of magnitude.  For scenarios with
unsteady flow, an accurate spatial ensemble estimate requires
averaging over multiple simulations.

We will use two sets of units in this paper: 1) \emph{system} units
that assume that the mass and length of the entire bound system along
with the gravitational coupling constant is unity; and 2) \emph{Milky
  Way} units that scale the model to our Galaxy using physical
units--kiloparsecs, years, and km/s.  When quantities are quoted
without units, they refer to \emph{system units} and when quantities
are physical units, they refer to \emph{Milky Way} units.

\begin{figure}
  \centering
  \subfigure[initial conditions]{
    \includegraphics[width=0.48\textwidth]{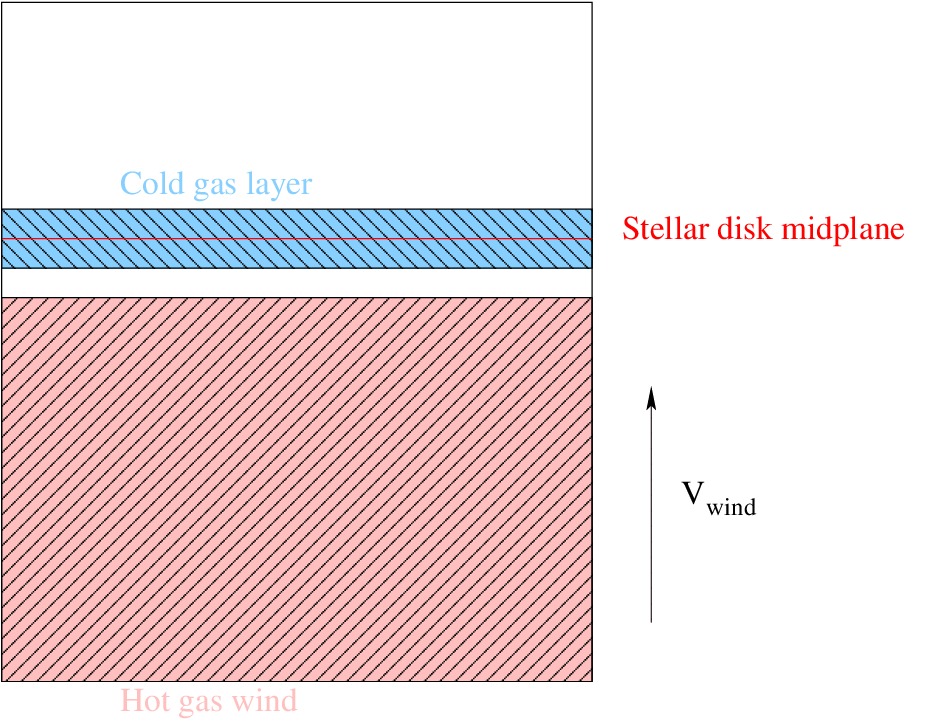}
    \label{fig:shockslab_sub1}
  }
  \subfigure[the wind-disk interaction]{
    \includegraphics[width=0.48\textwidth]{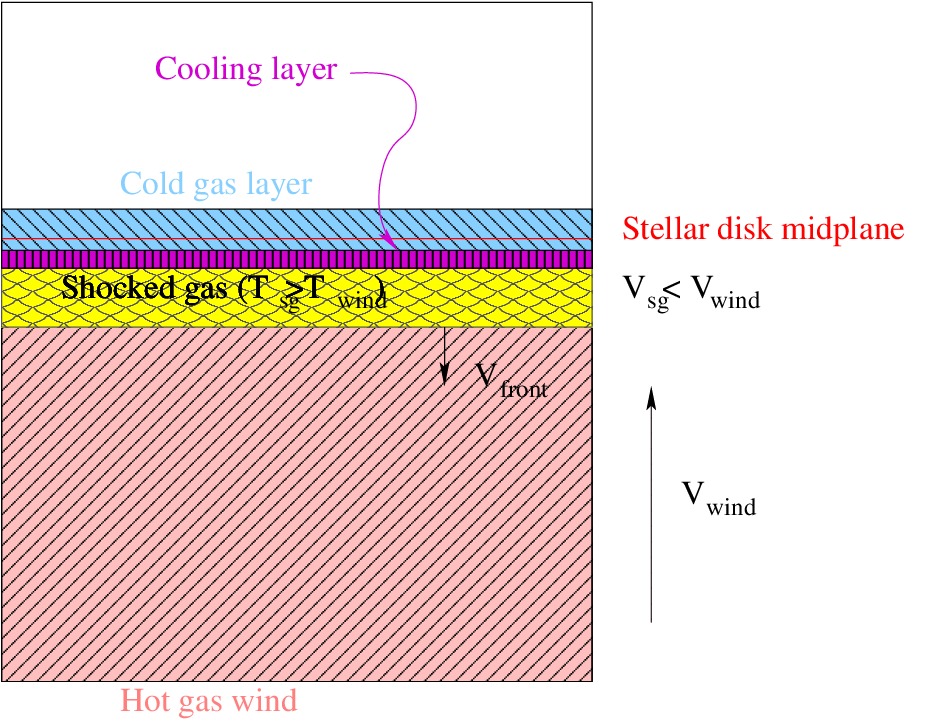}
    \label{fig:shockslab_sub2}
  }
  \caption{Description of the initial conditions
    \subref{fig:shockslab_sub1} and the evolving hot-wind--cold-disk
    interaction \subref{fig:shockslab_sub2} in the frame comoving with
    the disk.  The cold and hot gas is indicated by blue and red
    shading respectively.  In addition, the post-shock gas is shaded
    yellow and an enhanced cooling layer at the hot-cold interface is
    shaded purple.}
  \label{fig:shockslab}
\end{figure}

\begin{figure*}
  \centering
  \mbox{
    \includegraphics[width=0.5\textwidth]{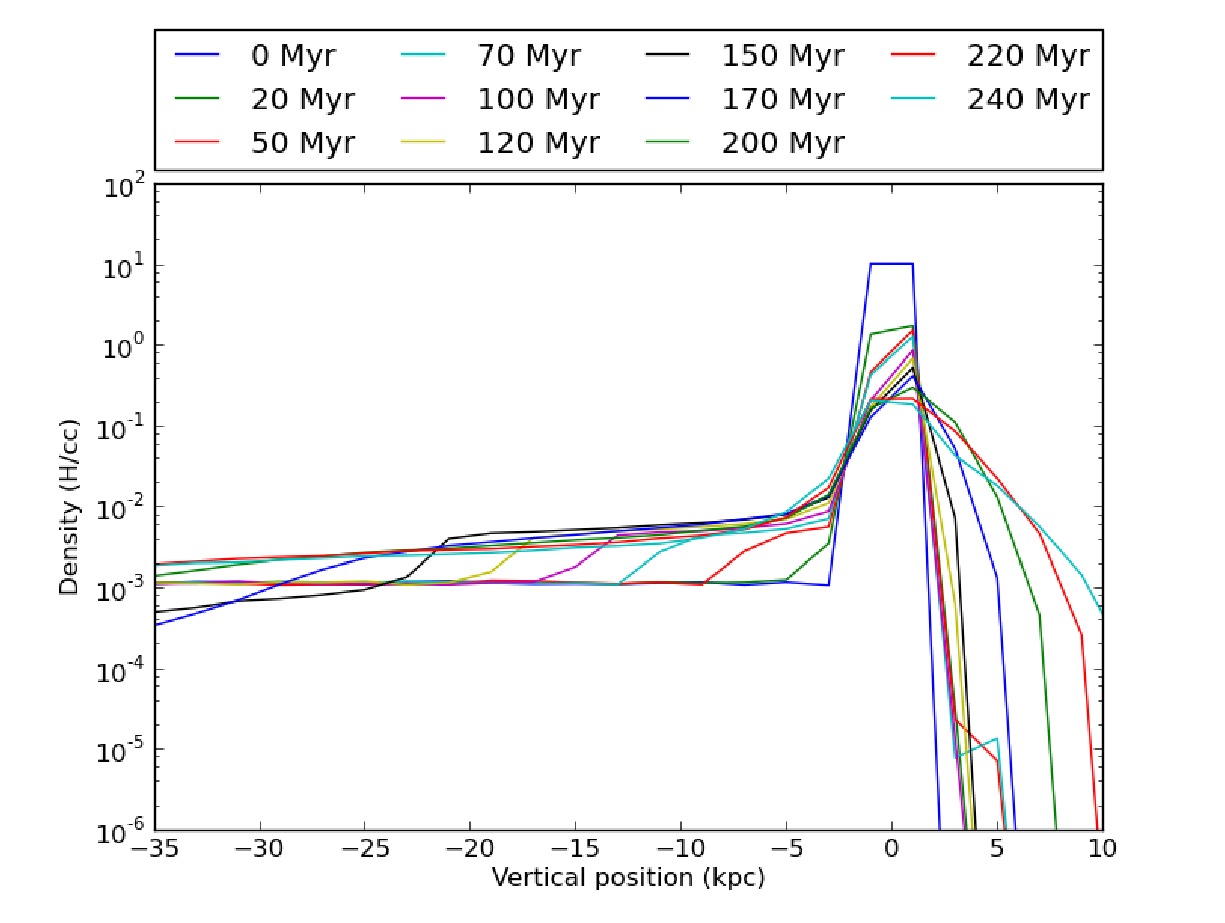}
    \includegraphics[width=0.5\textwidth]{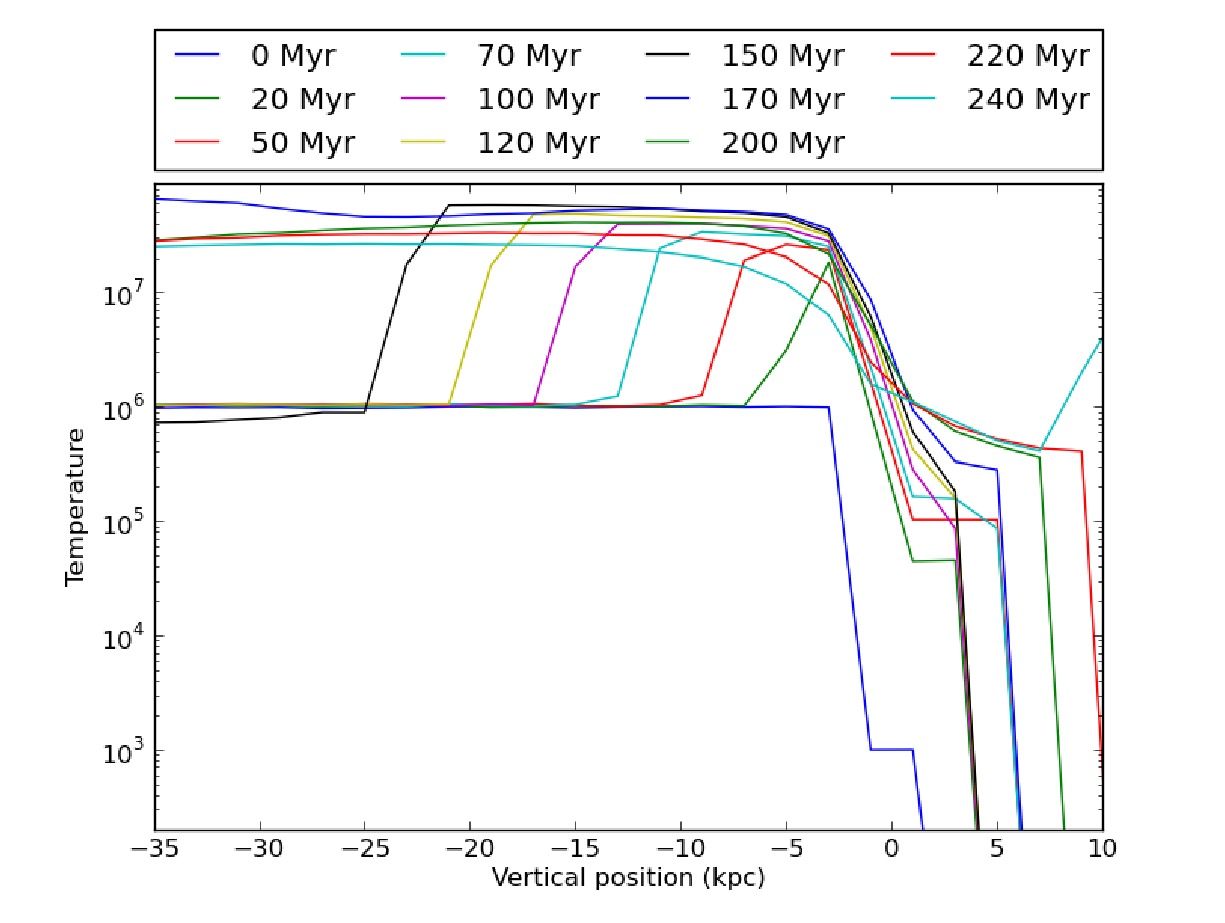}
  }
  \caption{Evolution of the gas density (left) and temperature (right) 
	for hot-wind--cold-disk interaction without cooling centred on the
    midplane of the stellar disk, $z_{centre}(t)$.  The horizontal
    offset from the stellar midplane is shown in kiloparsec and the
    density scale is in H atoms per cubic centimetre (H/cc).}
  \label{fig:slabtrace_nocool}
\end{figure*}

\begin{figure*}
  \centering
  \mbox{
    \includegraphics[width=0.5\textwidth]{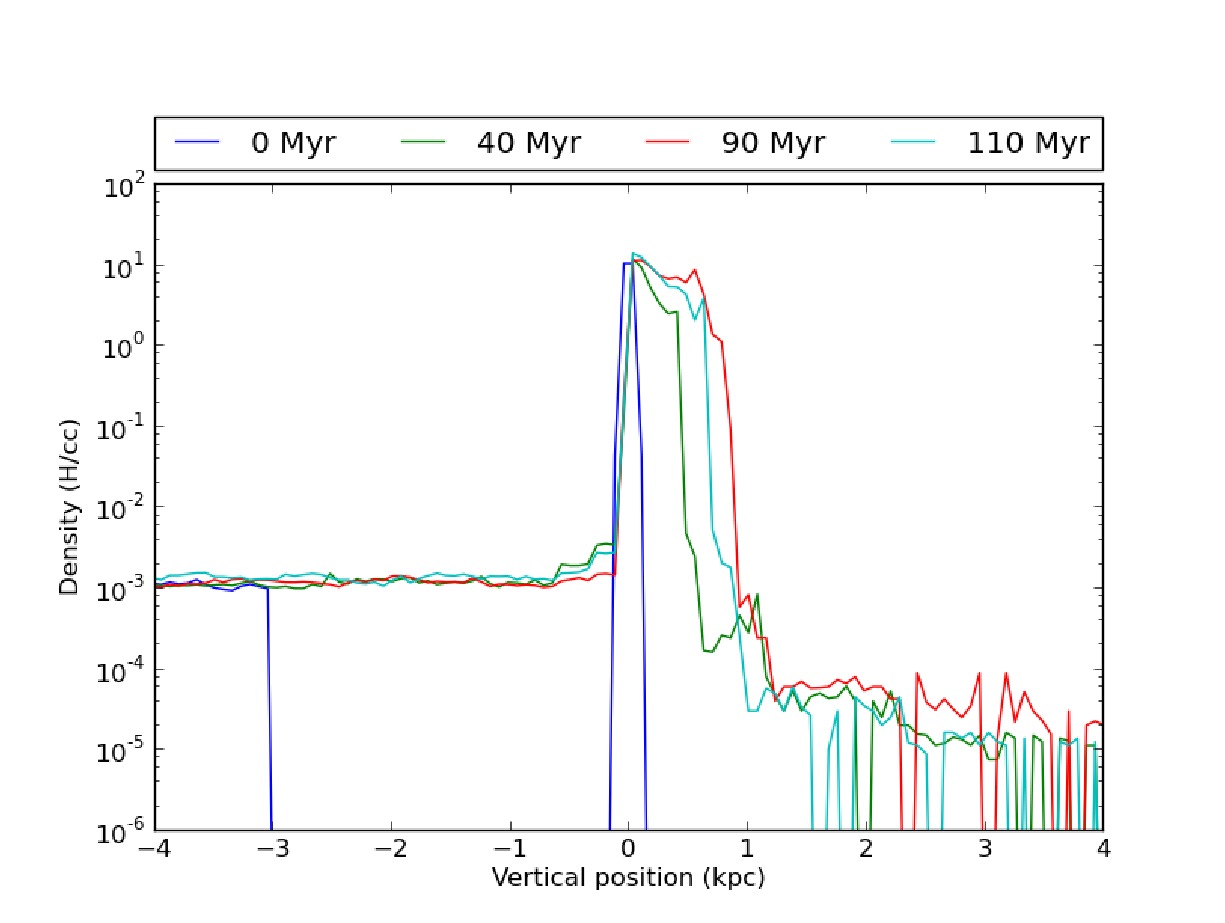}
    \includegraphics[width=0.5\textwidth]{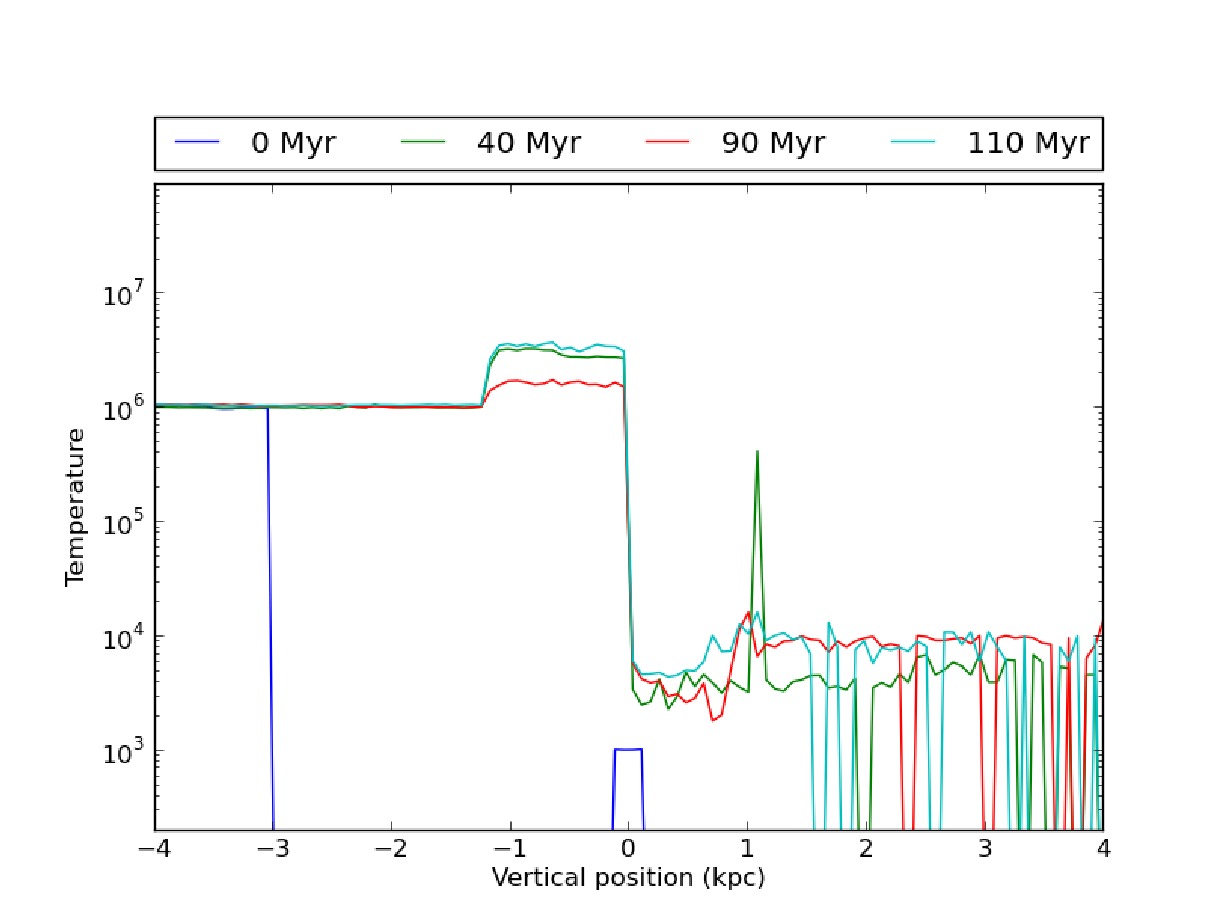}
  }
  \caption{Evolution of the density (left) and temperature (right) for
    hot-wind--cold-disk interaction, as described in
    Fig. \protect{\ref{fig:slabtrace}}, but \emph{with} atomic and plasma cooling.  The
    horizontal scale as been zoomed to the much smaller region
    containing the response.  The jagged distribution at $z\gta
    1$\,kpc owes to binning variance caused by the very small mass of
    gas above the disk. }
  \label{fig:slabtrace}
\end{figure*}

\subsection{Slab model: simple experiment}
\label{sec:simple_slab}

The first experiment is that of a one-dimensional slab of stars whose
gravity confines a cool gas layer.  This was designed to explicit test
the Gunn-Gott criterion in its natural domain of validity: a wind
impinging on a cold gas layer that is gravitationally confined to a
plane-parallel slab.  The stars have the classic isothermal
distribution $\rho(z)\propto\mbox{sech}^2([z - z_{centre}(t)]/h)$ with
scale height $h$ and this distribution is fixed for all time.
However, the centre of slab, $z_{centre}$, varies to conserve total
linear momentum for the entire system.  The gas is initialised in 
isothermal equilibrium
with the confining slab with $T=1000$ K and mid plane density of $n=1$
atom/cc.  The isothermal solution results in a gas scale height of
approximately $0.3 h$. The boundary conditions are periodic in the
plane of the slab ($x$ and $y$ directions) and vacuum in the vertical
($z$) direction.  The simulation begins with a hot homogeneous wind
($T=10^6{}$ K, $n=0.001$ atoms/cc) filling the region $z<-10h$ with a
mean vertical velocity chosen to exceed the Gunn-Gott criterion by a
factor of three.  As described above, all gas components may cool by
atomic and plasma cooling mechanisms following the standard LTE cooling
computation.  See Figure \ref{fig:shockslab_sub1} for a schematic
description.

Much to my surprise, the cold gas does not escape the disk potential!
Rather the evolutionary scenario is as follows.  The gas shocks as it
approaches the disk.  The shock heats the gas, and the upstream
compression wave propagates back into the original wind.  When cooling
is included, this shock stalls.  The velocity of the turbulent, denser
cooling layer of the post-shock hot gas is lower than $v_{wind}$ and
the pressure begins to displace the cold layer from mid plane.  The
momentum from the wind is transferred to the bulk ISM through a
cascade of primary and secondary inelastic scattering events. See
Figure \ref{fig:shockslab_sub2} for the location of these regimes.
Furthermore, the hot gas begins to pressurise the cold gas, exciting
two-phase instabilities that give rise to temperature and density
`fingers' as the $10^6$ K gas displaces the volume originally occupied
by the cold gas (not shown here).  We will see in Section
\ref{sec:gal_hard} that this production of multiphase features is
enhanced by rotational shear.  The total momentum is conserved but the
slowed, compressed flow downstream of the shock implies that the wind
does not transverse the disk on a ballistic crossing time.  Rather,
the slowed flow of hot gas increases the effective interaction time
and allows the stellar disk to respond dynamically to the displaced
cold disk.  Then, the hot wind's momentum is transferred to the
stellar disk gravitationally.  So in the end, the cold gas layer is
not lost but rather the entire disk, gas and stars together, is
accelerated, albeit weakly.

Figures \ref{fig:slabtrace_nocool} and \ref{fig:slabtrace} show the
density and temperature state perpendicular to the disk plane with
time.  We begin with the evolution of density and temperature without
gas cooling (Fig. \ref{fig:slabtrace_nocool}).  The curves at time 0
are the initial conditions.  Without cooling, the energy deposited by
the wind is sufficient to heat the initially cold gas layer to $10^6$
K in 200 Myr.  In this case, the shock forms and propagates upwind
without stalling.  The post shock wind gas increases its density
(temperature) by approximately a factor of 4 (10).  The dense
initially cold gas is able to radiate a large fraction of deposited
energy.  After approximately 200 Myr, the now heated low-density disk
component begins to flow downstream.  Although this result agrees with
the Gunn-Gott prediction, the mechanism is different: evaporation
rather than impulsive momentum transfer.

The addition of radiative cooling qualitatively changes the
ram-pressure response as shown in Figure \ref{fig:slabtrace}). Three
bins cover the initially cold disk (red).  On impact, the cold disk
heats to approximately $10^4$ K and thereafter radiates much of the
wind-deposited kinetic energy.  The shock forms, begins to
propagates upwind, and stalls.  The ISM reaches a new approximate
thermal equilibrium between the kinetic heating and radiative cooling
at approximately $10^4$ K.  The gas density decreases while the scale
height of the initially colder gas increases.  The distribution of the
disk gas is skewed relative to the midplane determined by the stellar
component but very little gas escapes.

In summary, the scenario described in Section \ref{sec:gunn_gott}
failed to describe the flow because the momentum transfer is
\emph{not} in the impulse limit as derivation of equation
(\ref{eq:ggc}) assumes.  Owing to the shock, the post shock gas acts
as a slowing moving piston.  This gives time for the stellar disk to
respond gravitationally to the displacement of the cold disk from the
stellar mid plane.  As a result, the initially cold gas layer is
decelerated by the stellar disk causing the momentum from the wind to
be transferred to the disk, and in the long run the entire system,
although the disk-halo interaction occurs on a much larger timescale
than in this simulation.  Cooling is an essential ingredient that
prevents the cold gas disk from evaporating.  Without cooling, the
disk gas is eventually evaporated, however with cooling, very little
gas escapes.

\subsection{Galaxy model: harder experiment}
\label{sec:gal_hard}

\begin{figure*}
  \mbox{
    \includegraphics[width=0.5\textwidth]{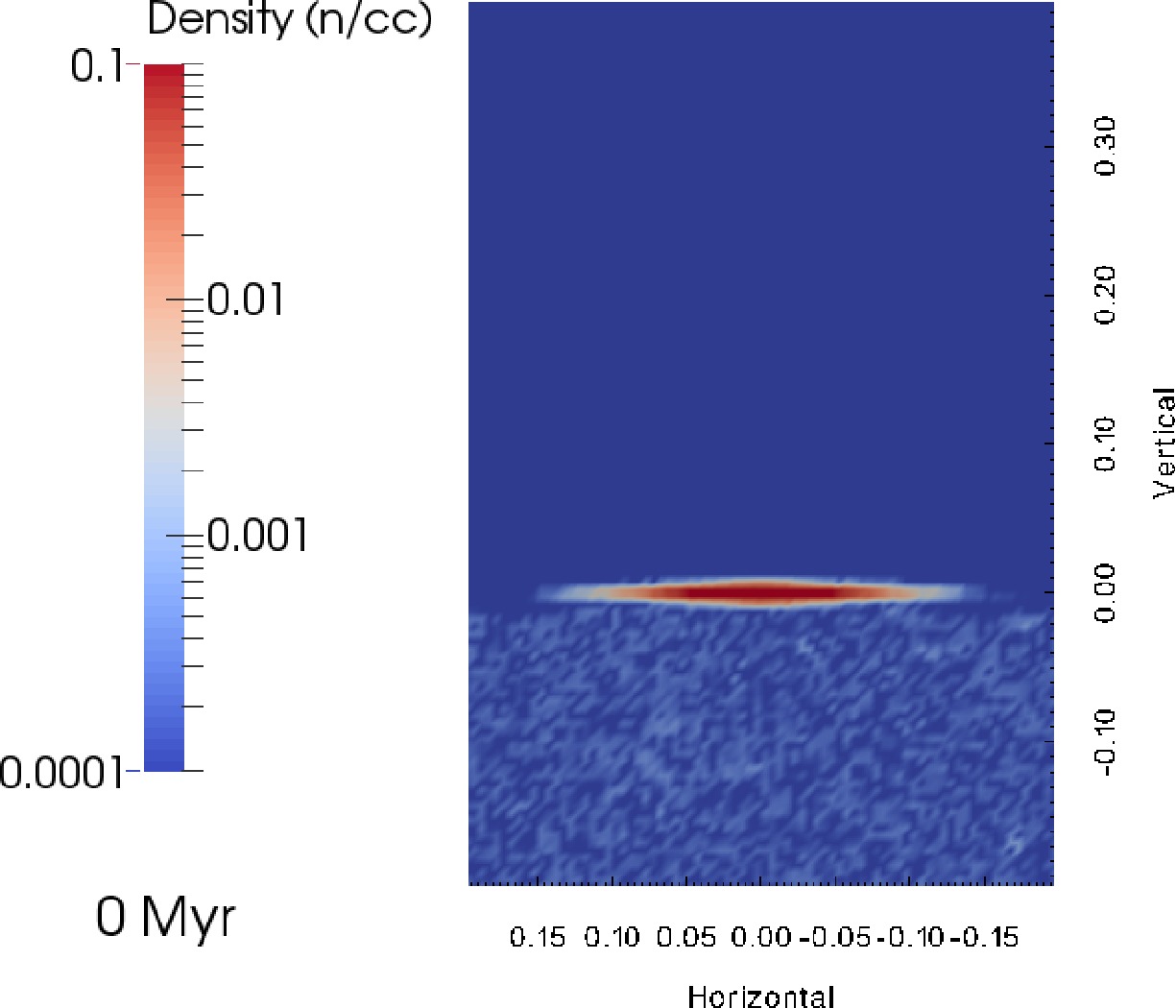}
    \includegraphics[width=0.5\textwidth]{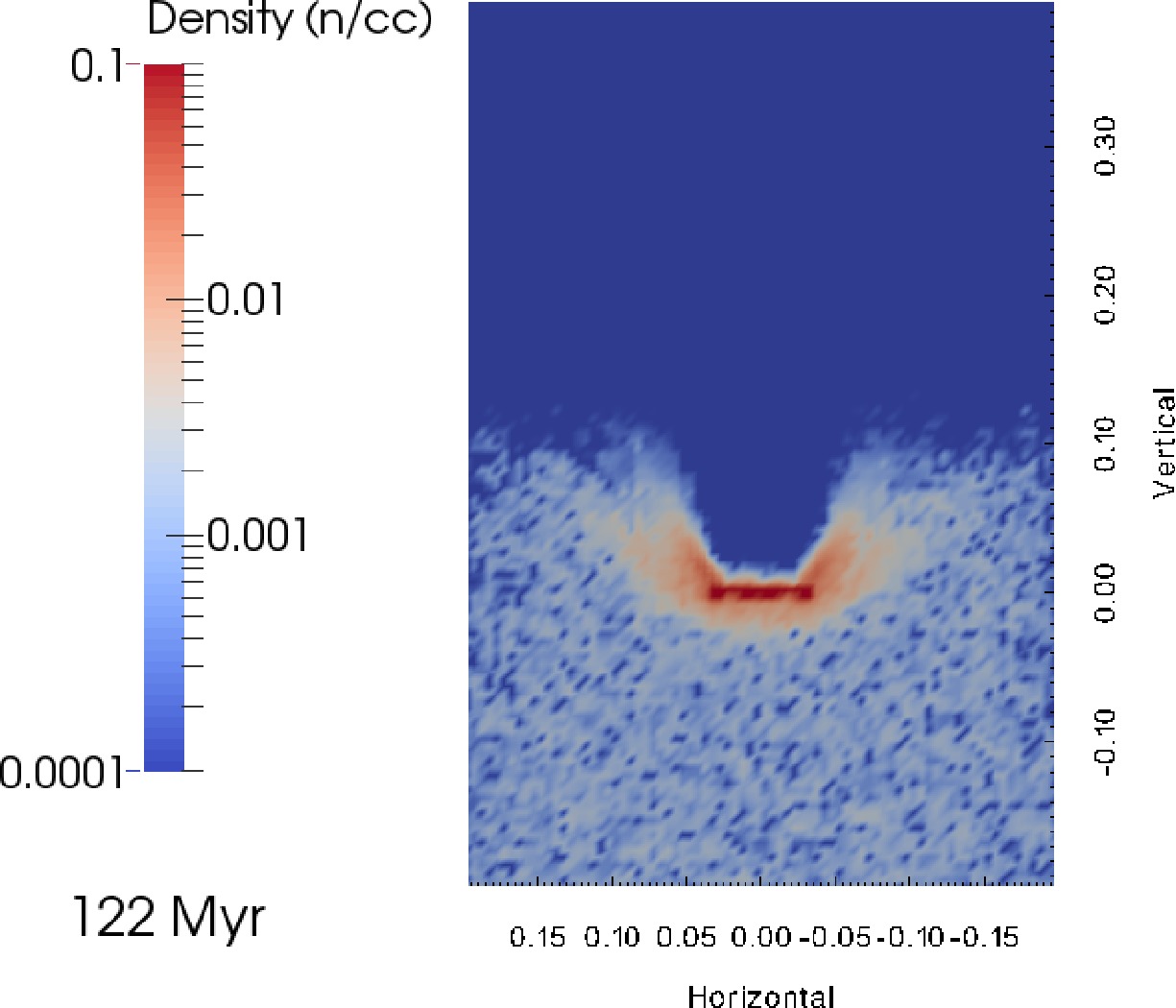}
  }
  \par\bigskip
  \mbox{
    \includegraphics[width=0.5\textwidth]{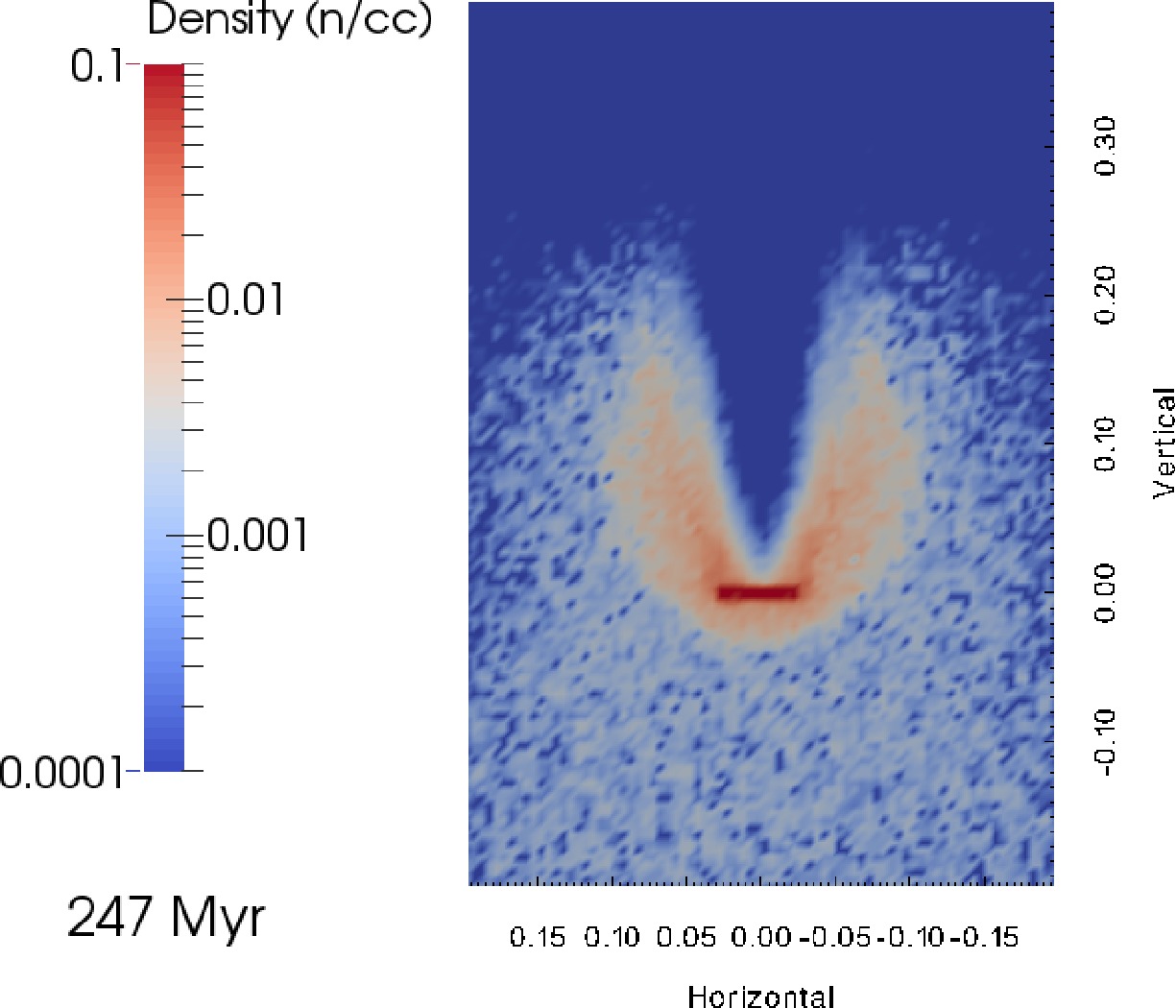}
    \includegraphics[width=0.5\textwidth]{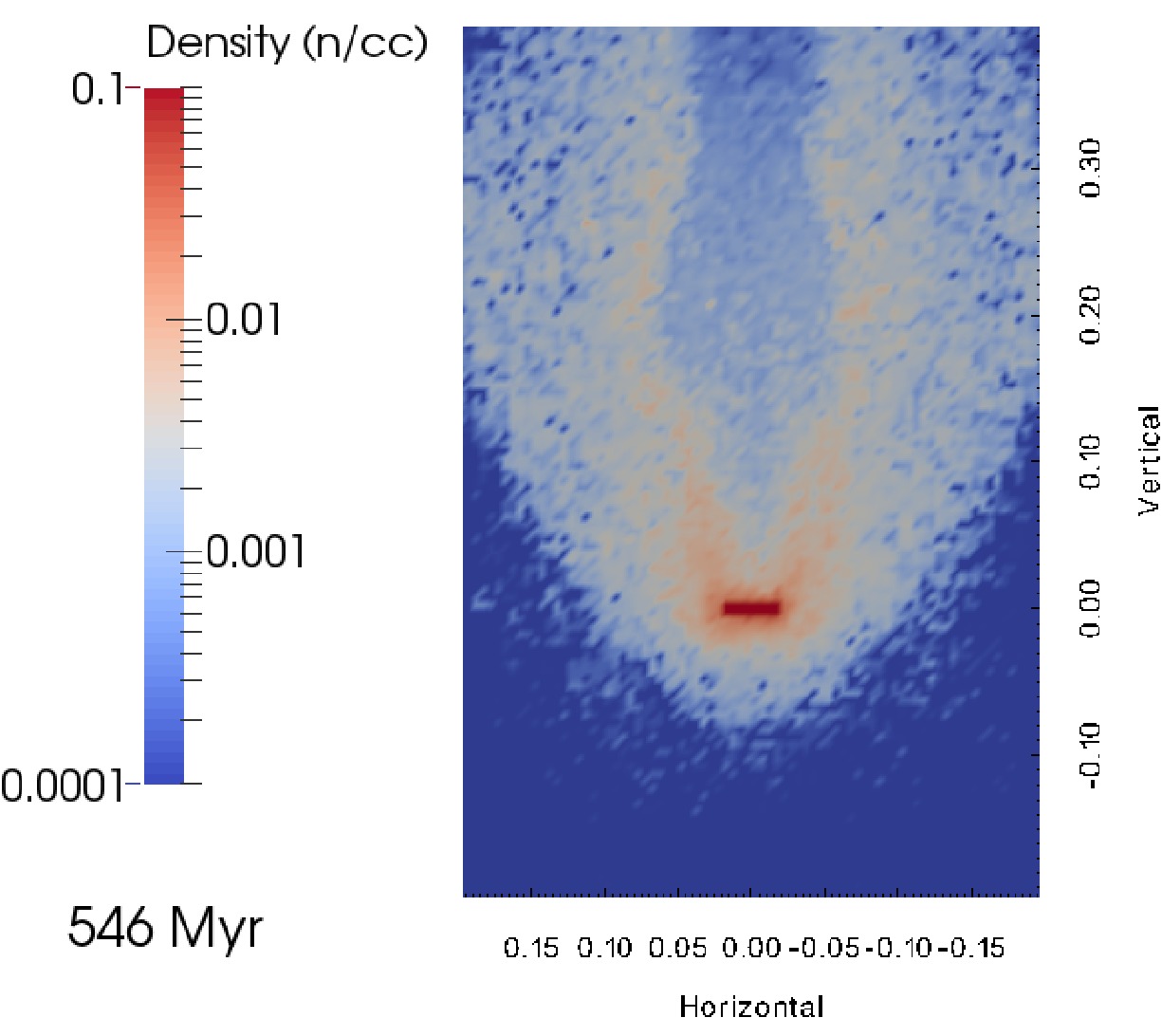}
  }
  \caption{Edge-on view of the combined hot-wind--cold-disk gas
    evolution.  Each panel describes the density in units of
    hydrogen atoms per cubic centimetre.  One length unit is
    approximately 300 kpc scaled to the Milky Way galaxy.  The
    elapsed time is inset in the lower right-hand side of each
    panel.  The flow is cylindrically symmetric on large scales.
    The cold, dense disk is seen between $z=-0.01$ and $0.01$ (wind
    direction) and $x=-0.04$ and $0.04$ (transverse direction).  The
    notches of lower density in the disk gas are due to strong
    two-phase instabilities.  }
  \label{fig:shockedgaldens}
\end{figure*}

\begin{figure*}
  \mbox{
    \includegraphics[width=0.5\textwidth]{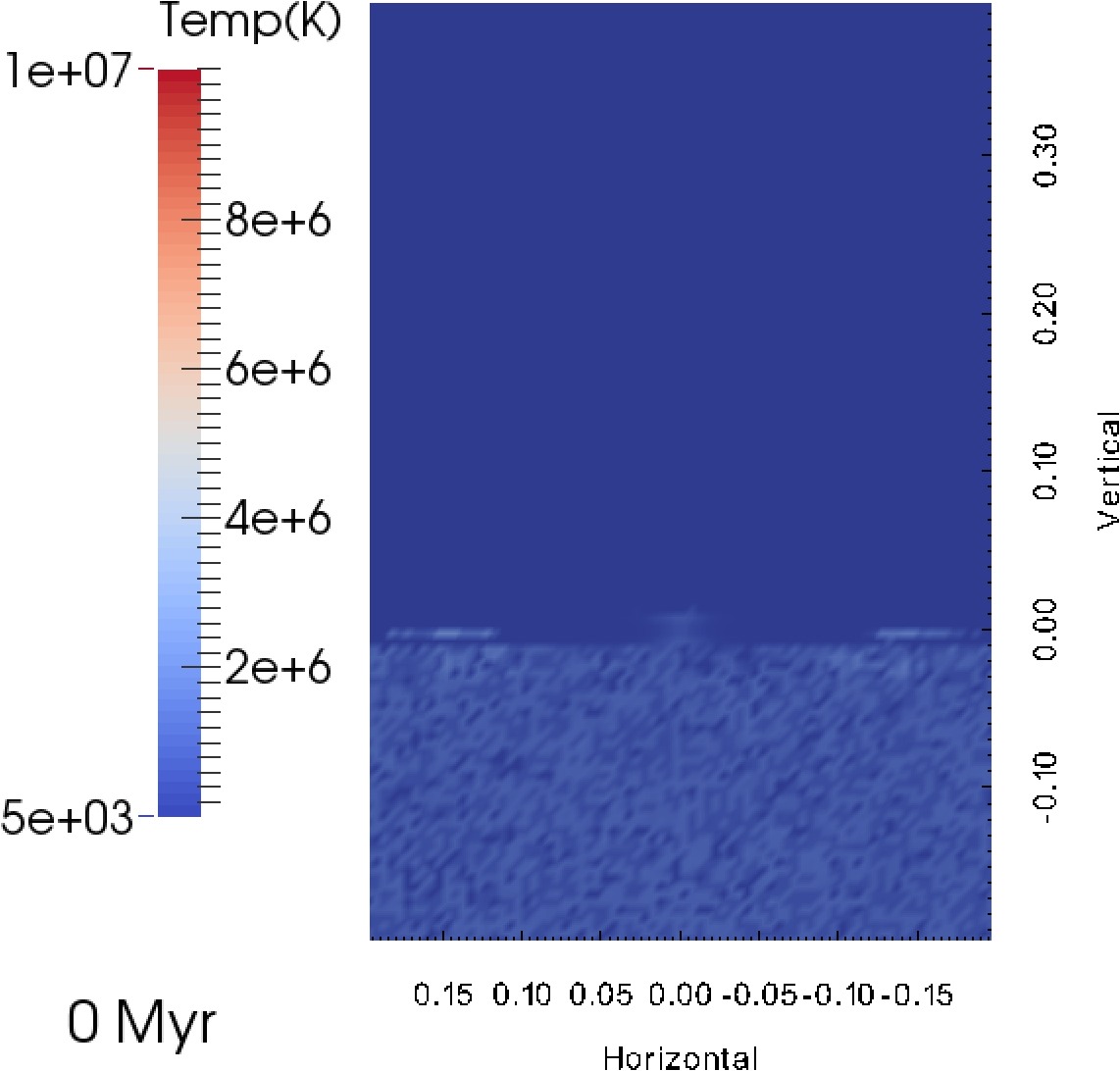}
    \includegraphics[width=0.5\textwidth]{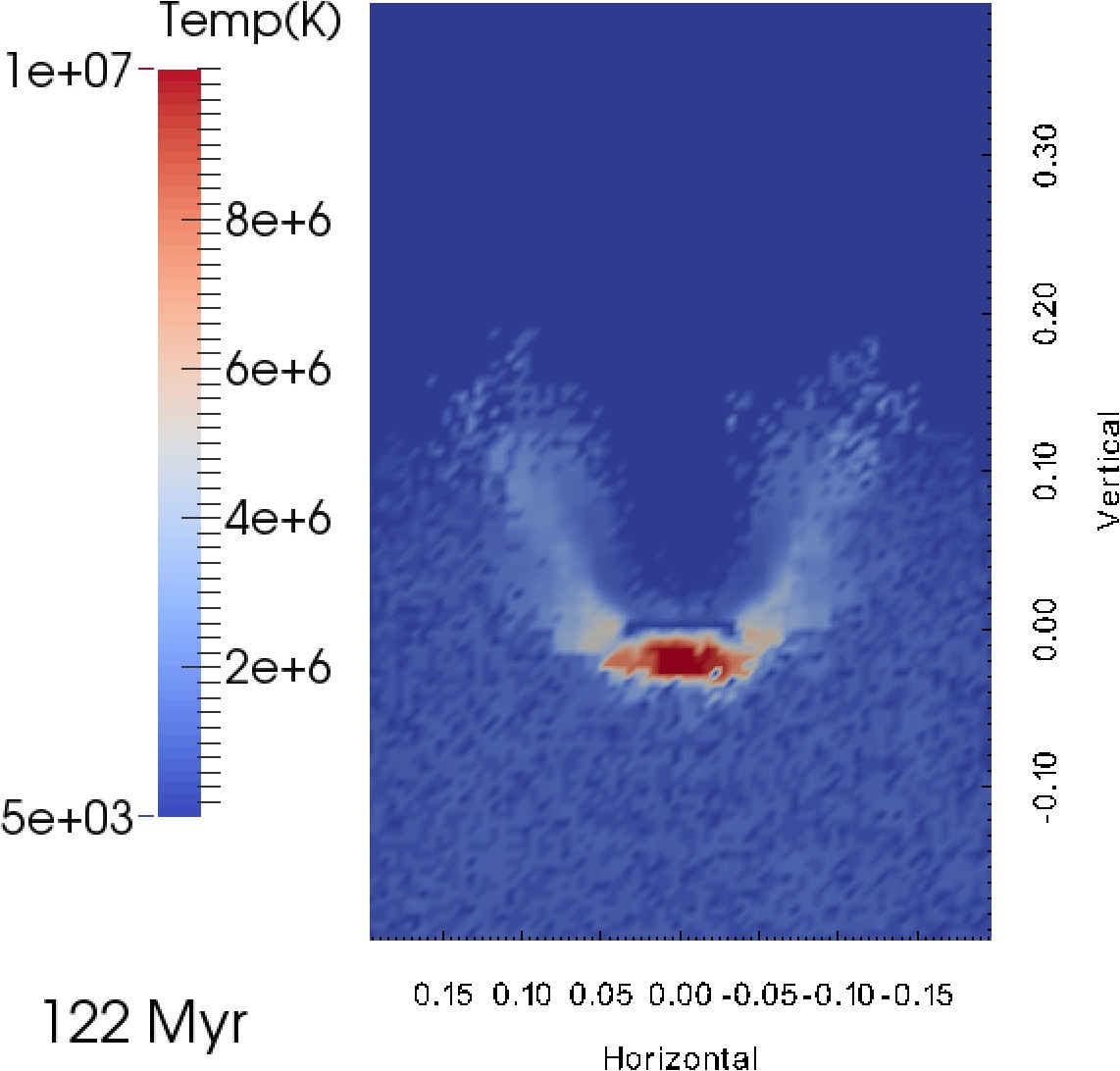}
  }
  \par\bigskip
  \mbox{
    \includegraphics[width=0.5\textwidth]{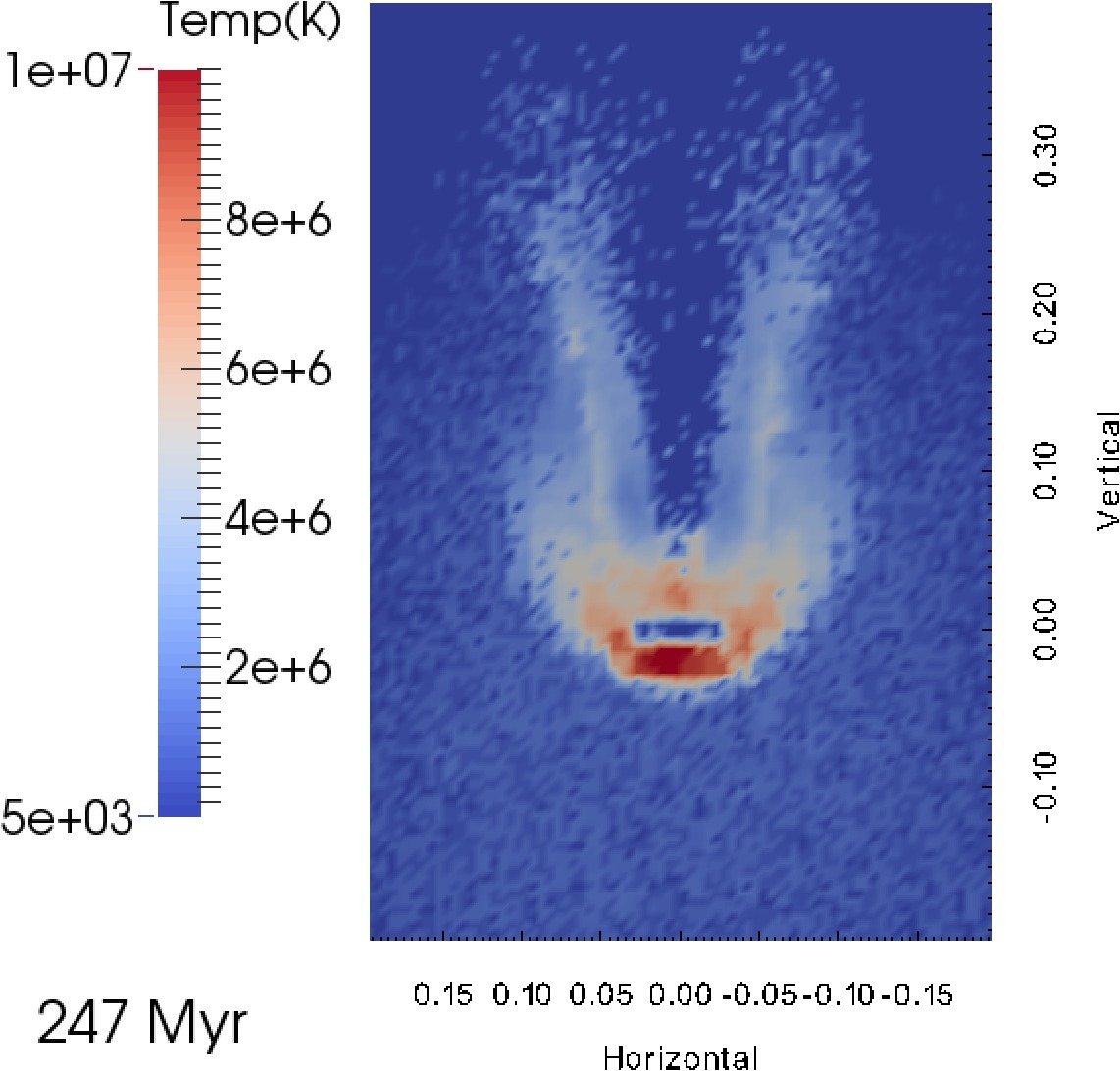}
    \includegraphics[width=0.5\textwidth]{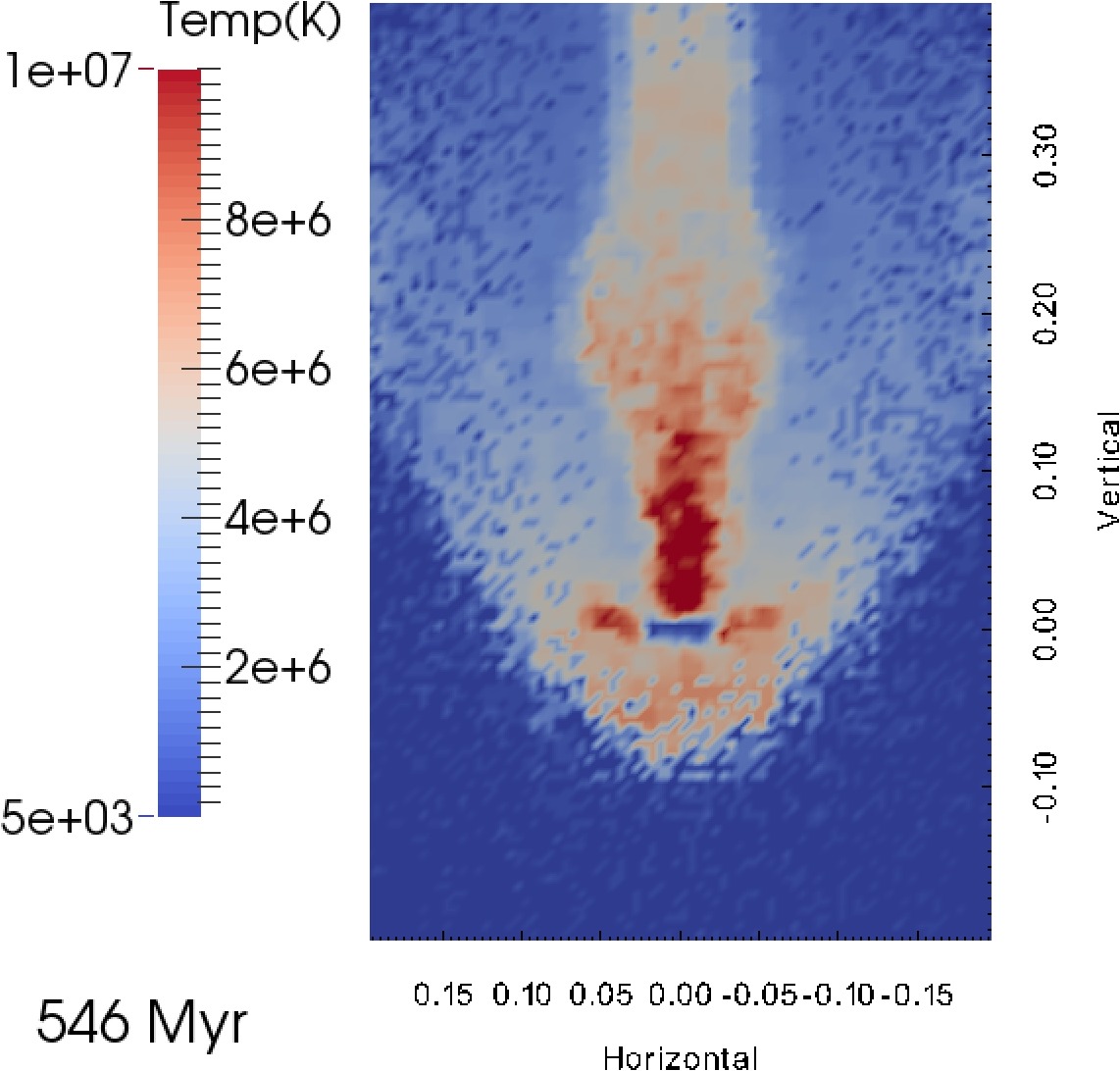}
  }
  \caption{Edge-on view of the combined hot-wind--cold-disk gas
    evolution.  Each panel describes the temperature in degrees K.
    The cold gas disk and the hot Mach disk are clearly seen at $z=0$
    and $z=-0.02$, respectively.  The hot 'plume' visible between
    $z=0.01$ and $0.06$ is actually a hot low-density sheath supported
    against radial collapse by rotation.  }
  \label{fig:shockedgaltemp}
\end{figure*}

\begin{figure}
  \centering
  \subfigure[isodensity surface]{
    \includegraphics[width=0.49\textwidth]{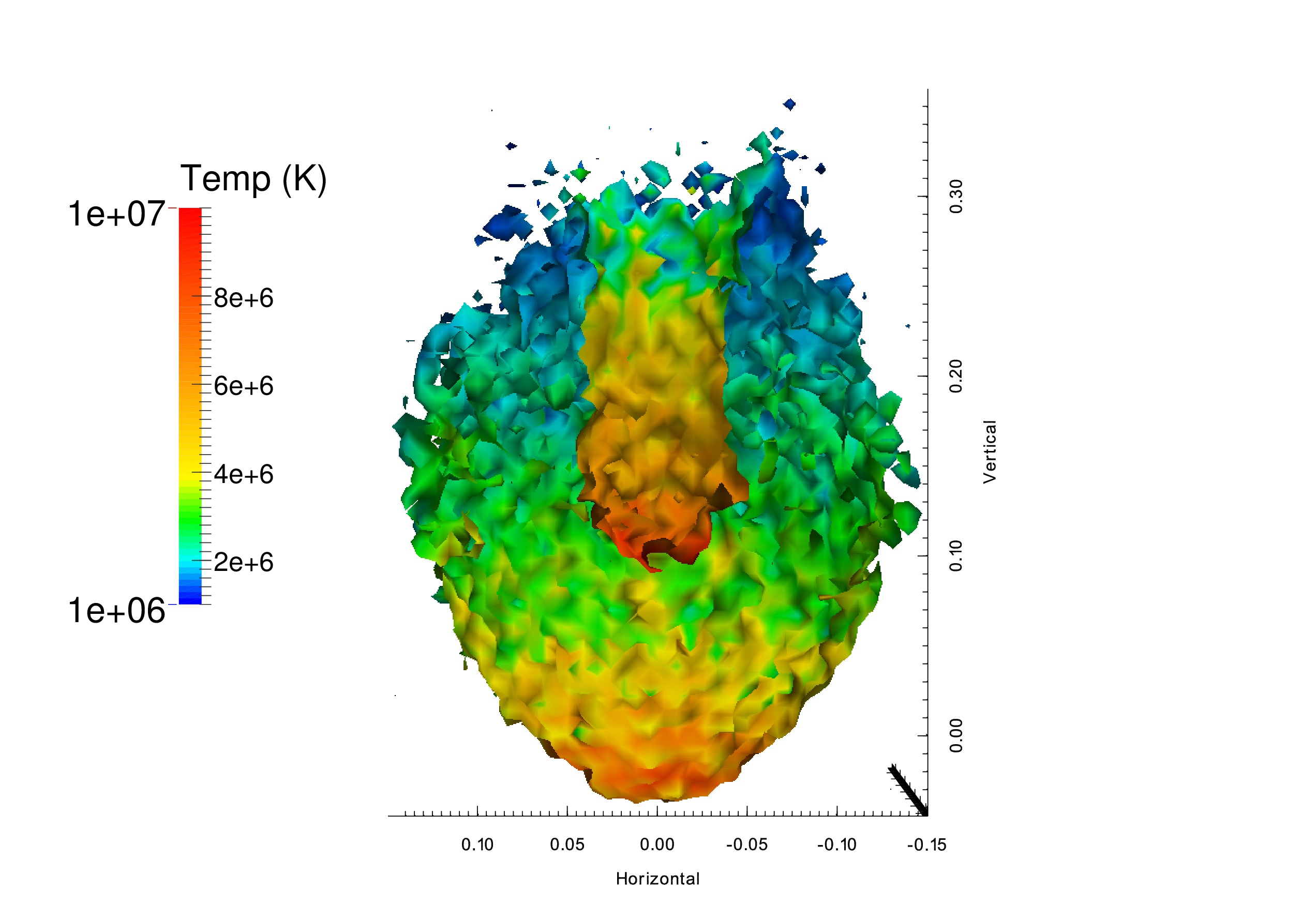}
  }
  \subfigure[streamlines]{
    \includegraphics[width=0.49\textwidth]{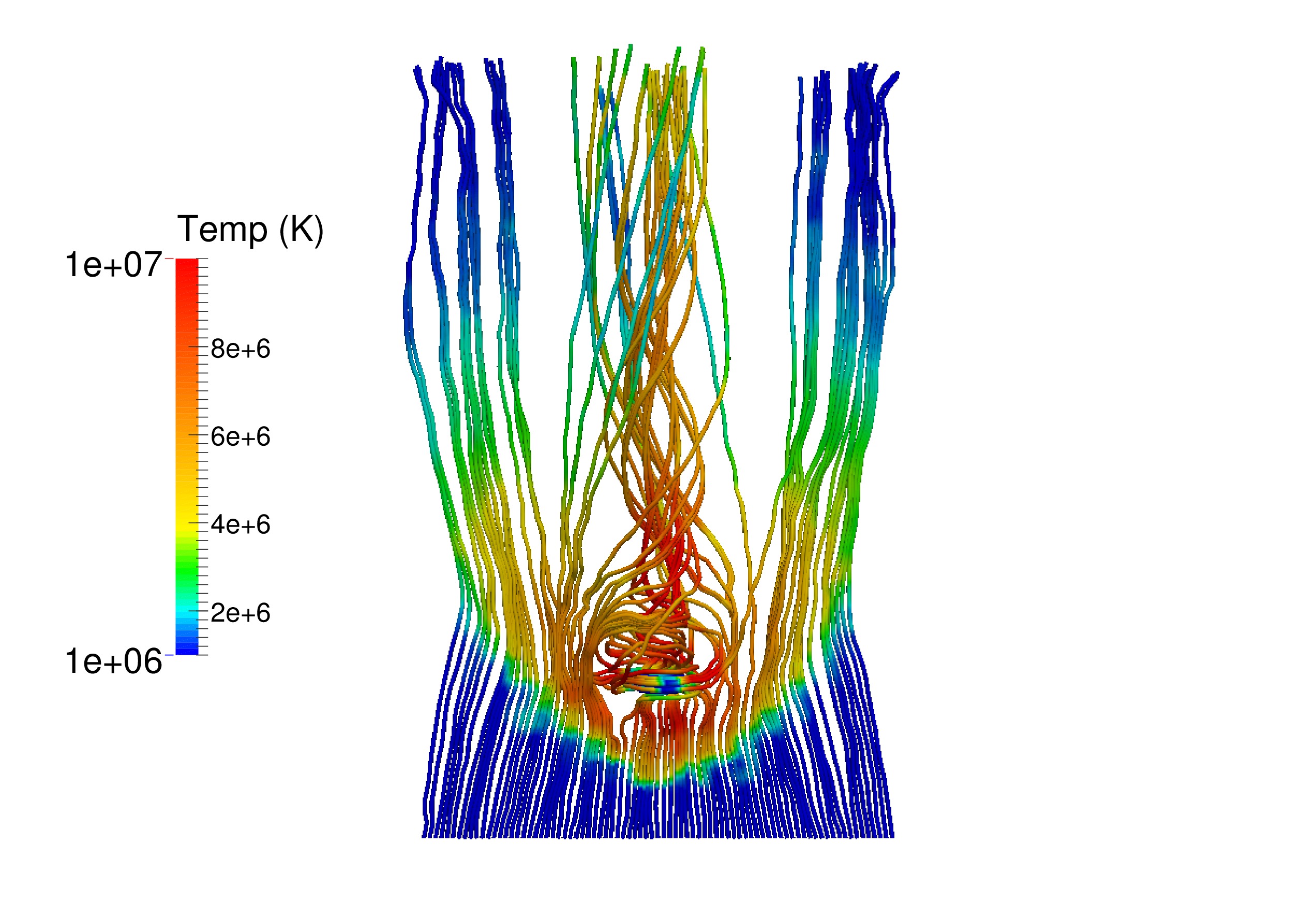}
  }
  \caption{Density and streamline rendering of the gas flow at the
    same point shown in the last frame of
    Fig. \protect{\ref{fig:shockedgaldens}}.  Panel (a) show a
    level-set surface corresponding to gas density of 0.003
    particles/cc.  This surface is approximately cylindrically
    symmetric, but has been cut by a vertical plane passing through
    the axis of symmetric to reveal its structure: a hollow sheath
    with the shape of a hornet's nest beginning with the bow shock at
    the bottom which a large narrow downward dimple corresponding to
    the inner sheath.  The surface is colour coded by temperature,
    showing that the inner sheath is nearly $10^7$ K as it leaves the
    disk.  Panel (b) shows the stream lines colour coded by the
    temperature scale.  The inner sheath in bounded by spiraling
    stream lines whose rotation comes from the angular momentum of the
    cold disk gas.  }
  \label{fig:gas_struct}
\end{figure}

\begin{figure*}
  \centering
  \includegraphics[width=0.8\textwidth]{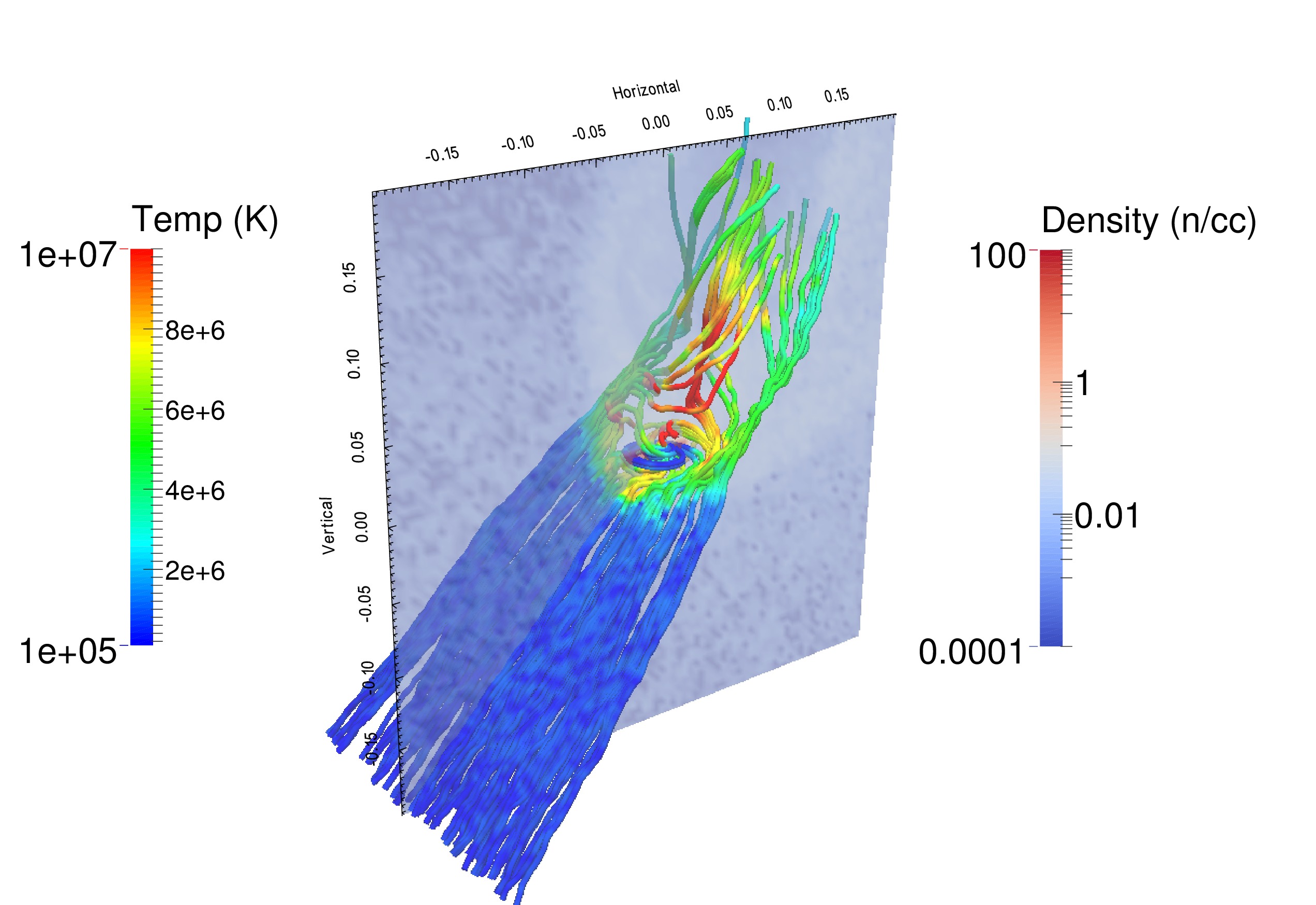}
  \caption{Streamlines showing gas flow for a hot wind with an impact
    angle of 45 degrees with a cold gas disk at 373 Myr after impact.
    The gas density of the flow in units of atoms/cc is shown on the
    translucent plane.  The overall morphology of the flow is
    qualitatively similar to that for face-on impact.}
  \label{fig:oblique_stream}
\end{figure*}

The first experiment was intentionally idealised and designed to check
the Gutt-Gott mechanism in the simplest conceivable scenario.
Intuitively, the inhomogeneity in the galaxy density profile will
change both the downstream flow patterns and possibly entrain cold
gas.  How will this change the simple scenario above from
Section \ref{sec:simple_slab}?

To test this, we use the more realistic equilibrium galaxy model with
the same wind column as in the previous experiment.  The wind is only
distributed in the vicinity of extended disk to best use computational
resources as described in Section \ref{sec:ics}.  Wind particles are 
not created during the simulation, and this results in wind--disk interaction 
event of limited duration; when most of the wind particles have reached 
the disk, the interaction is over.  To prevent the adiabatic
expansion of the wind column perpendicular to the flow, the gas is
subjected to periodic boundary conditions in the two perpendicular
directions.  The duration of the simulation then is limited by either
(i) the complete passage of the wind column past the galaxy or (ii)
the wrapping of the flow transverse to the wind direction, which ever
happens first.  The vertical extent of the wind column is chosen,
empirically, so that these two conditions obtain at approximately the
same time.  For simulations here, this is approximately 500 Myr after
the wind impact, scaled to a galaxy of Milky-Way size.

Vertical slices containing the geometric centre of the disk through
the gas density and temperature fields constructed from the ensemble
values of the DSMC particles are shown in Figures
\ref{fig:shockedgaldens} and \ref{fig:shockedgaltemp}.  The initial
condition, just before the vertically moving wind reaches the disk is
depicted in the top-left panels of Figures \ref{fig:shockedgaldens}
and \ref{fig:shockedgaltemp}. The disk plane is initially at $z=0$.
As in the simple slab model, the wind gas shocks as it reaches the
galaxy.  Initially, the momentum of the wind atoms is transferred
directly to the cold gas.  However, after the shock forms, the
momentum from the incoming wind is transferred by the shock to the
slower, compressed downstream flow.  The heated post-shock gas slows
and piles up, increasing its density and driving a compression wave
upstream into the wind.  After approximately 100 Myr (the lower-right
panels in Figs. \ref{fig:shockedgaldens} and
\ref{fig:shockedgaltemp}), a well-defined Mach disk has formed with a
temperature of close to $10^7$ K, offset approximately 0.02 units
below the disk plane (approx. 6 kpc, scaled to the Milky Way).
Meanwhile, the hot wind begins to flow around the disk, heating and
ablating the outer initially cold disk gas. Most of the disk gas is
\emph{ablated} at approximately 3 disk scale lengths.  The temperature
of the ablated gas is approximately $8\times10^6$ K.  This gas cools
to approximately $4\times10^6$ K as it expands and forms a bell-shaped
sheath.

Between 150 to 300 Myr after the initial impact, the gas continues to
flow around and ablate the cold disk.  The post-encounter wind mixed
with the ablated gas forms a hot sheath.  Owing to the angular
momentum of disk gas ablated from approximately 3 disk scale lengths,
the sheath has positive angular momentum with respect to the
cylindrical axis.  The unshocked wind, flowing around the galaxy,
pressurises the sheath but the inward motion toward the symmetry axis
is halted by its angular momentum barrier.  This results in a
relatively empty hot central core (final panel of
Fig. \ref{fig:shockedgaldens}).

By 300 Myr, the hot post-shock gas at impact parameters smaller than 2
scale lengths has begun to push itself through the cold disk.  This
has three consequences: 1) the additional pressure provided by the hot
gas drives a strong two-phase cooling instability; 2) as the hot gas
is pushed through the hot holes created by the instability, it
receives some of the angular momentum of the cold gas.  This flow is
now partly rotationally supported against inward radial motion,
creating a second inner sheath.  Inside this sheath, there is only a
hot tenuous gas with a much lower gas density than that of the initial
wind (by approximately an order of magnitude); 3) the shear created by
the vertical oozing of the hot gas through the now clumpy cold-gas
layer excites a Kelvin-Helmholtz instability in the disk plane (this
will be described in detail below).

The fully developed wind-cold disk interaction at the end of the
simulation is described in the final panel of Figures
\ref{fig:shockedgaldens} and \ref{fig:shockedgaltemp}.  The final
panel of each of these figures describes the flow at the end of the
simulation ($t\approx 550$ Myr).  The hollow inner sheath begins to
propagate upward in a spiral owing to the net angular momentum
obtained from the disk.  The inner sheath is very hot ($T\approx10^7$
K) but with low density ($\lta 10^3$/cc).  Figure \ref{fig:gas_struct}
further illustrates the state of the gas at 550 Myr.  The upstream gas
is gravitationally focused toward the galaxy giving the isodensity
contours a hornet's nest shape. Panel (a) shows a rendered surface at
a constant density of 0.002 atoms/cc colour-coded by gas temperature in
degrees Kelvin.  At the bottom of the surface sits the Mach disk at a
temperature of $8\times10^6$ K.  The projected rectangular feature
between $z=0.1$ and $0.3$, $x=-0.05$ and $0.05$ is the inner boundary
of the sheath; you are seeing a cut through a low-density tube.  Panel
(b) depicts stream lines integrated through the gridded volume used to
compute the ensemble field quantities, also coded by gas temperature.
The location of the shock front is clearly seen by the sharp jump in
the temperature of the ambient wind gas from $10^6$ K to
$T>4\times10^6$ K.  After impact, the stream lines flare outward again
and the gas cools a small amount.  Approaching the inner galaxy, the
hot gas attempts to push through the disk.  In doing so, the hot gas
gains some angular momentum and the cool gas loses some angular
momentum.  In the end, some net angular momentum is lost to the hot
gas and the resulting spin can be seen in the streamlines.  A small
fraction of the low-density gas squeezes through the multiphase disk
at small radii and small specific angular momentum to form a funnel of
hot gas with $T\gta3\times10^6$ K.

This same numerical experiment has been repeated with a wind direction
of 45 degrees to the disk normal direction with very similar
results. Figure \ref{fig:oblique_stream} describes the resulting flow
with temperature-coded stream lines, as in Figure \ref{fig:gas_struct}
with a density field slice along the symmetry axis.  The stream lines
and density slice reveal the same inner and outer sheath morphology as
Figure \ref{fig:gas_struct}.  The cold disk response and the
angular momentum transport are also quite similar to the face-on case,
to be described below, so we will not consider the details of the
oblique-wind simulation any further.

Typical of the galaxies with strong ICM interactions described in
\citet{Chung.etal:2009}, NGC 4522 shows little H{\sc I} in the outer
disk.  Even more remarkable is the appearance of some gas above the
inner disk, consistent with our prediction that the hot gas will force
itself through the hot phase of the two-phase unstable disk during the
later phases of the interaction.  The extent of the neutral gas
provides a further interesting diagnostic for the ICM--ISM
interaction.  Our simulation predicts that ram-pressure stripped
hydrogen will not be observable in its neutral phase
(cf. Fig. \ref{fig:shockedgaltemp}) owing to the heat provided by the
shock and carried by the turbulent flow which rapidly heats the
initially cold gas in the outer galaxy.  There are some caveats: the
ICM density used here is larger by an order of magnitude and the wind
velocity used here is smaller by nearly an order of magnitude than
those values appropriate for Virgo.  Given the complexity of this
interaction, a direct prediction should be made with the appropriate
values.  In addition, because this simulation does not include
gravitation on small scales; it is possible that this simulation
misses small scale cold-blobs forming from cooling instabilities.
However, the pervasive shock heating suggests that H{\sc I} tails
detected in galaxies that also show evidence of ram-pressure
interactions \citep[e.g.][]{Vollmer.Huchtmeier:2007} have lost their
neutral gas through tidal stripping.  Nonetheless, the morphology of
gas in NGC 4522 is qualitatively similar to that seen in these
simulations and certainly appears to be an ongoing wind-disk
interaction as indicated by \citet{Vollmer.etal:2008a}.  It is
conceivable that both mass-loss mechanisms are occurring
simultaneously in NGC 4330 \citep{Abramson.etal:2011}.  The time scale
for survival of tidally-lost neutral gas in the hot turbulent wake of
the ICM--ISM interaction may be distinctly morphology dependent and
will require further work.  Future work using the full DSMC
collisional implementation will yield detailed radiative diagnostics
(see Section \ref{sec:summary}) which will help disentangle the
mechanisms.

\begin{figure}
  \begin{minipage}{0.475\textwidth}
    \includegraphics[width=\textwidth]{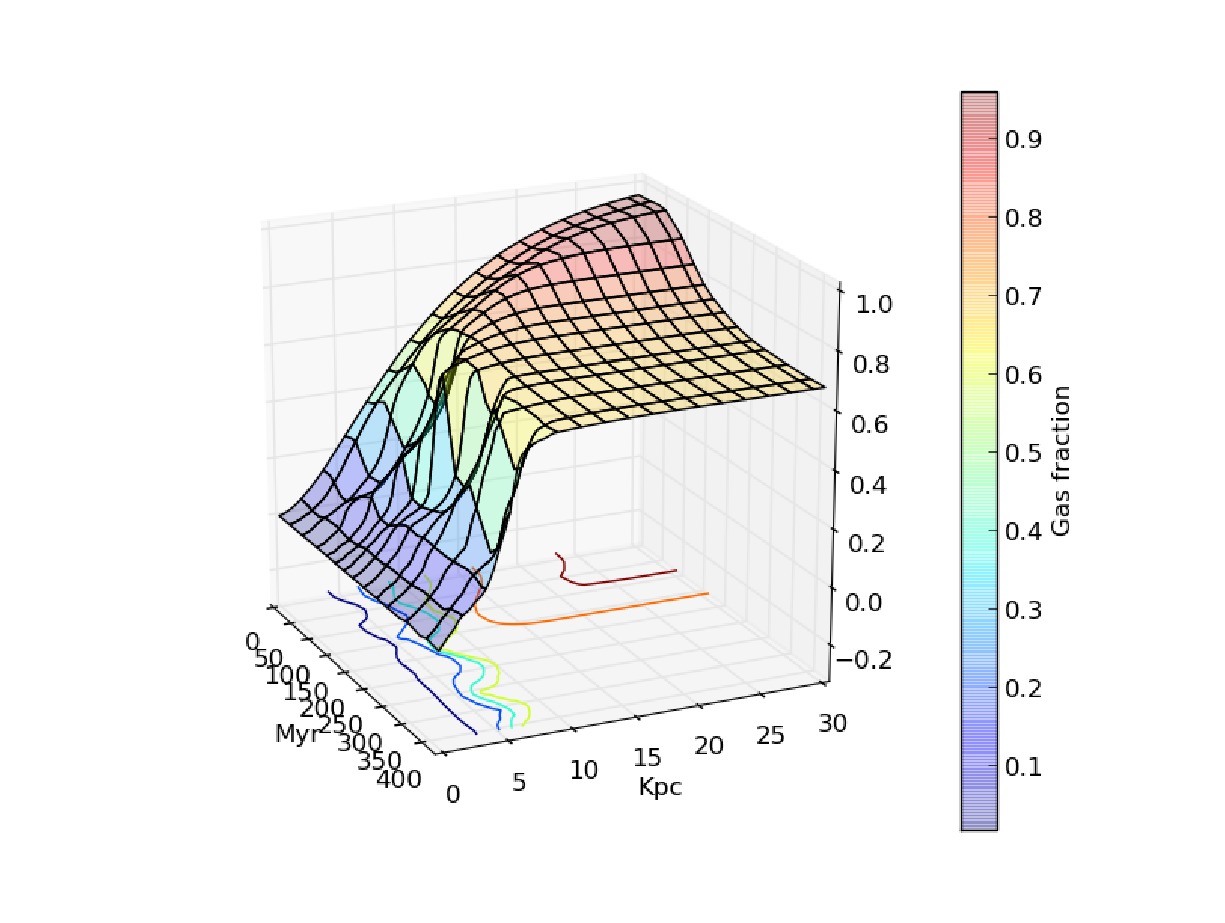}
    \caption{The cumulative fraction of gas in the disk with radius (in
      kpc) and time (in millions of years).  The lower plane has been
      dropped below zero to reveal the contour lines.}
    \label{fig:gas_fraction}
  \end{minipage}
  \hspace{0.05\textwidth}
  \begin{minipage}{0.475\textwidth}
    \includegraphics[width=\textwidth]{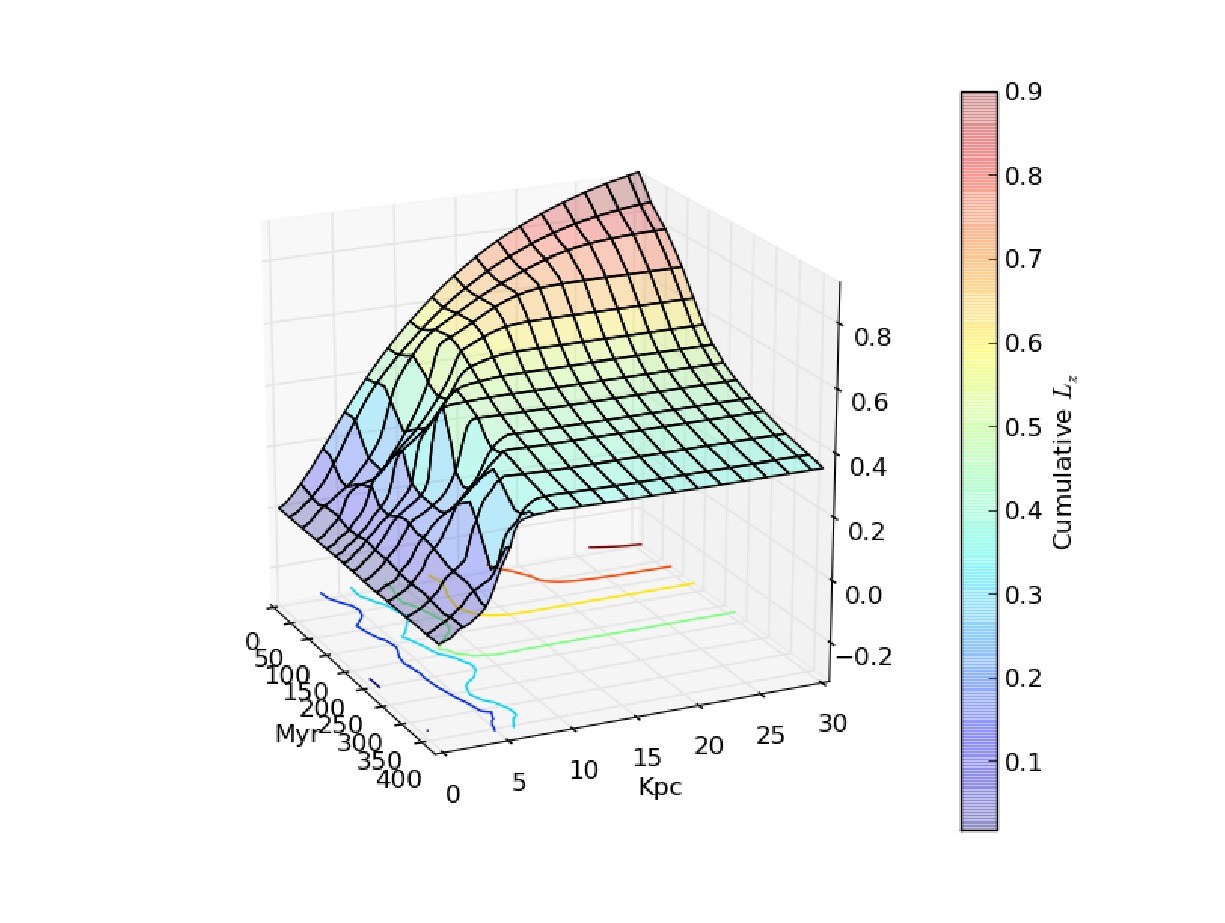}
    \caption{Similar to Fig. \protect{\ref{fig:gas_fraction}}, this
      figure shows the evolution of the cumulative angular momentum
      fraction of gas in the disk with time.  The contours on the base
      plane show fixed fractions for interpretive guidance.}
    \label{fig:gas_angmom}
  \end{minipage}
\end{figure}

Figure \ref{fig:gas_fraction} shows the azimuthally averaged
distribution of the disk's gas mass inside of a given radius (in kpc)
and as a function of time (in Myr).  We see that approximately $40\%$
of the initial gas mass is lost, and this is mostly from the outer
disk and in the first 300 Myr.  As seen in Figures
\ref{fig:shockedgaldens}--\ref{fig:shockedgaltemp}, the initially cold
gas in the outer disk is ablated and heated.  Although the gas
continues to be ablated at a low level, the loss rate is diminishes
rapidly after 300 Myr owing to both the increased density of the cold
gas after pressurisation and by the decrease in the scale length of
the cold gas after its sizeable loss of angular momentum.

Figure \ref{fig:gas_angmom} shows the evolution of the angular
momentum fraction with time.  This further emphasises that the disk
gas is peeled off from the outside in and demonstrates that the
overall gas content of the inner galaxy is largely unaffected. More
than $60\%$ of the original angular momentum content of the cold gas
disk is lost to the wind, the dominant fraction owing to the mass
loss.  The wiggles in the constant angular momentum fraction contours
are due to planar Kelvin-Helmholtz instabilities which we will
describe next.

\begin{figure*}
  \centering
  \mbox{
    \includegraphics[width=0.5\textwidth]{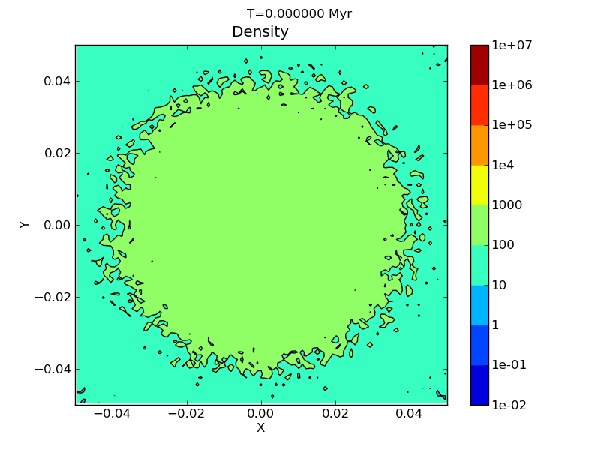}
    \includegraphics[width=0.5\textwidth]{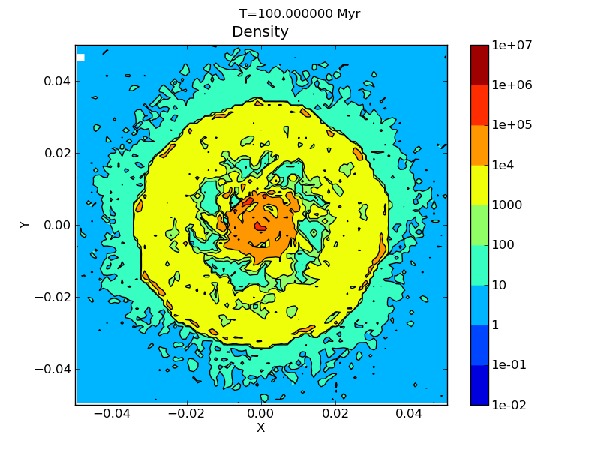}
  }
  \mbox{
    \includegraphics[width=0.5\textwidth]{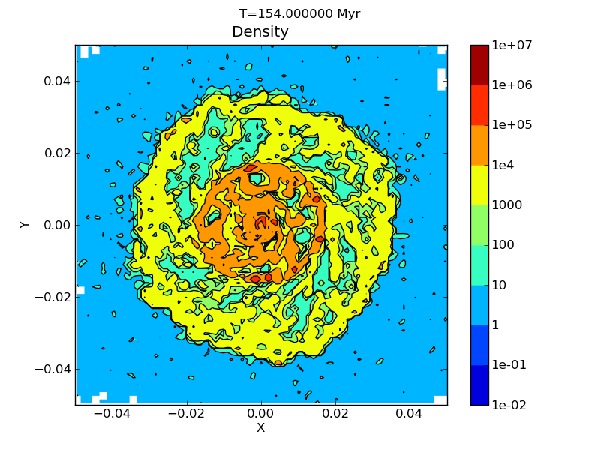}
    \includegraphics[width=0.5\textwidth]{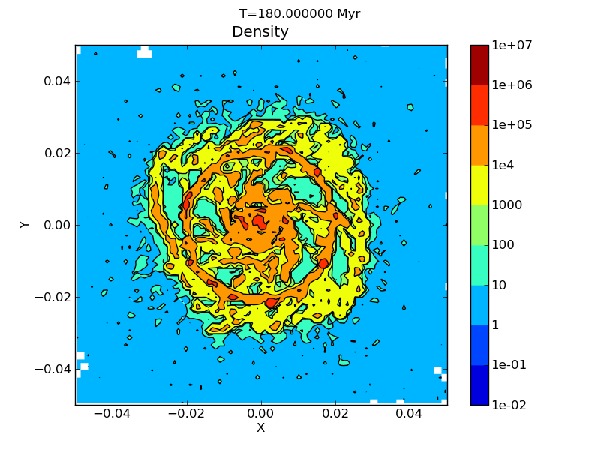}
  }
  \mbox{
    \includegraphics[width=0.5\textwidth]{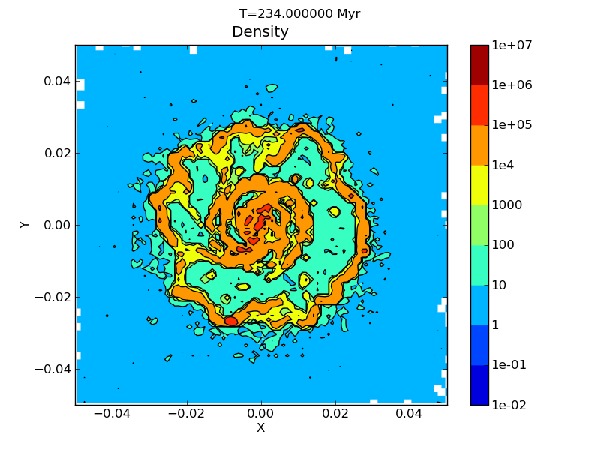}
    \includegraphics[width=0.5\textwidth]{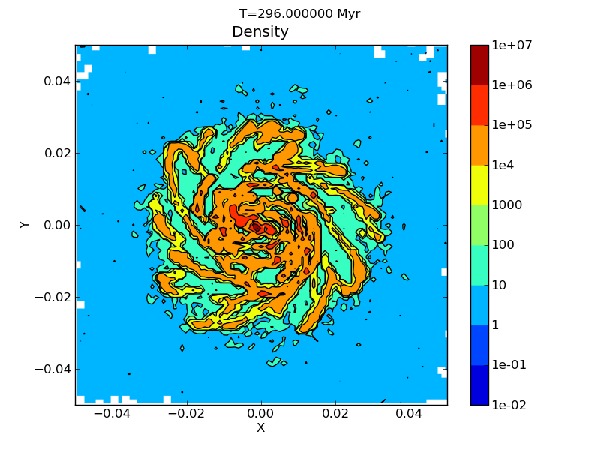}
  }
  \caption{Gas density as a function of time (in system units) at the
    disk mid plane.  System density units are $1.5\times10^{-3}
    \mbox{atoms}/\mbox{cm}^3$ and one scale length is 0.01 system
    length units on the $x$ and $y$ axes.  The time at the top of each
    frame is shown in millions of years (Myr).}
  \label{fig:face_on}
\end{figure*}

\begin{figure*}
  \includegraphics[width=0.49\textwidth]{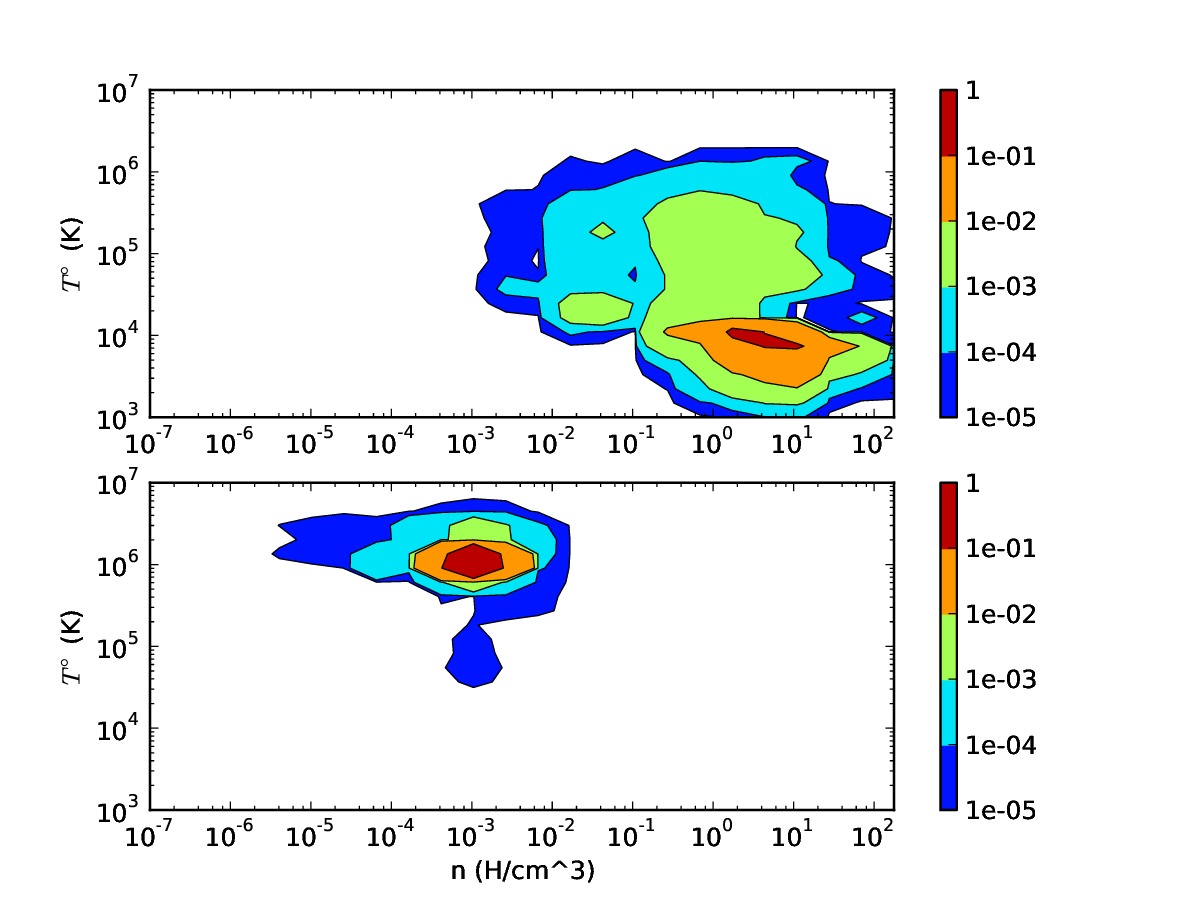}
  \includegraphics[width=0.49\textwidth]{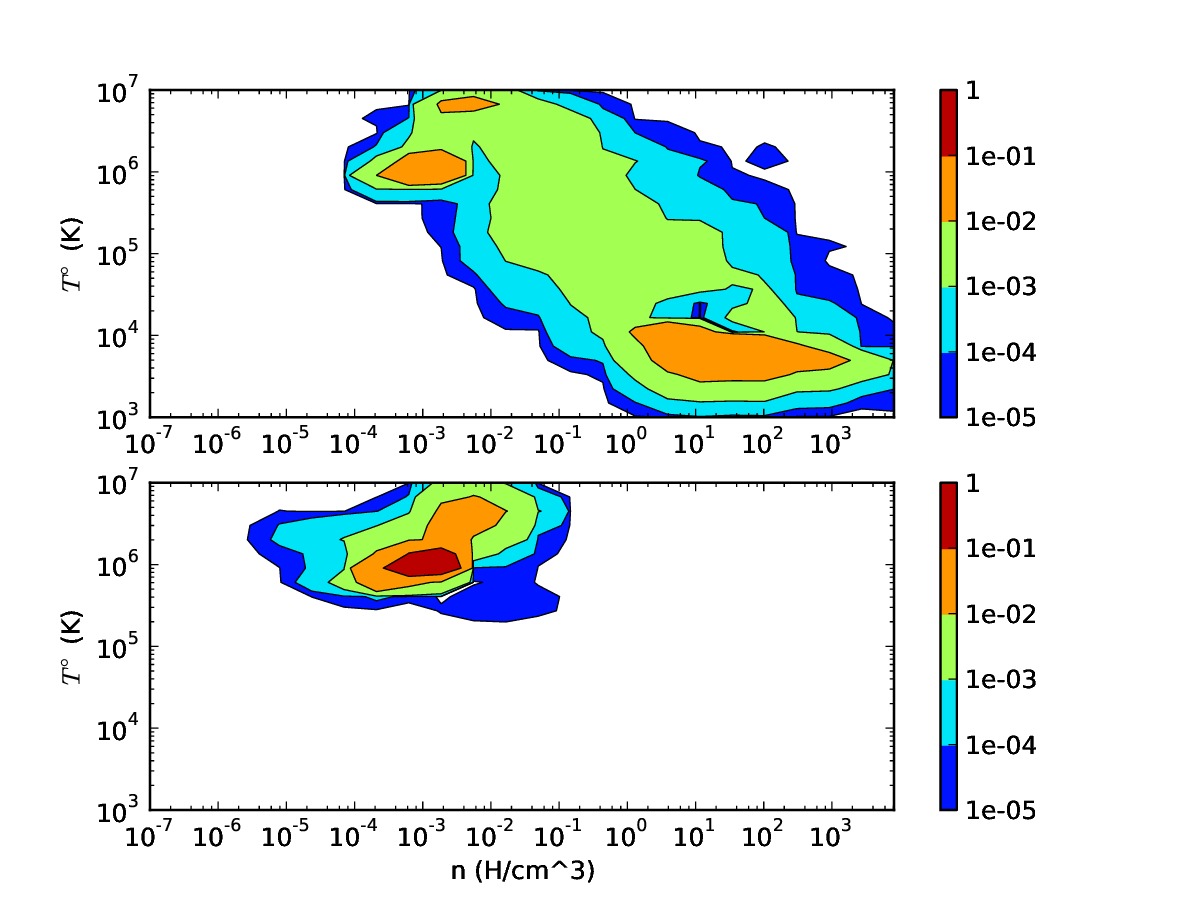}
  \caption{Density--temperature phase plane after 25 Myr (left column) 
    and after 200 Myr (right column)
    for the initially cold disk gas (top row) confined to the region
    of the stellar disk, and the initially hot wind gas (bottom row),
    filling the region initially below but later above and below the
    stellar disk.  The logarithmic colour scale denotes relative
    density on the phase place.  The initially cold gas develops a
    distinct hot tenuous phase through interaction with the hot wind
    (top row).  Similarly, the hot wind is both heated and cooled
    through its interaction with the cold disk gas (bottom
    row). \label{fig:density_temp1}}
\end{figure*}

\begin{figure*}
  \includegraphics[width=0.49\textwidth]{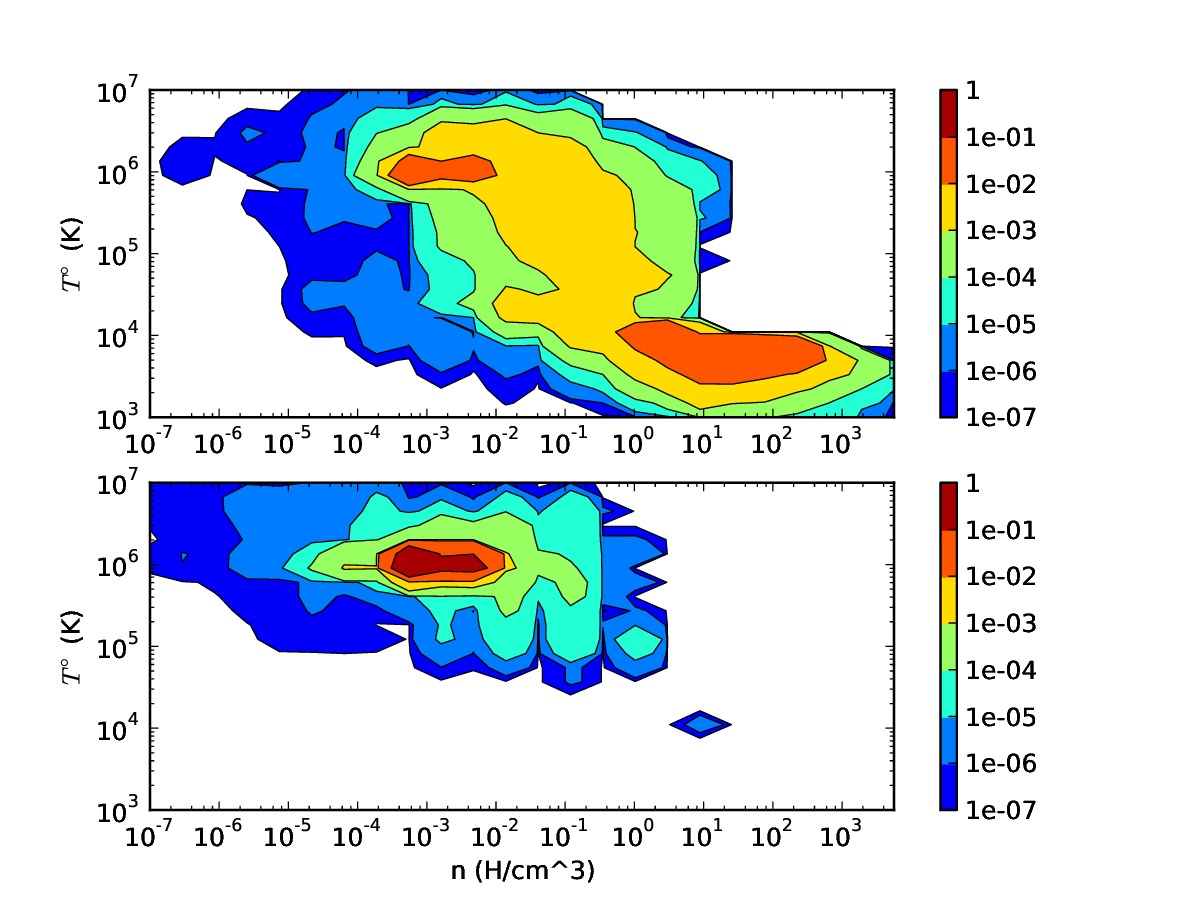}
  \includegraphics[width=0.49\textwidth]{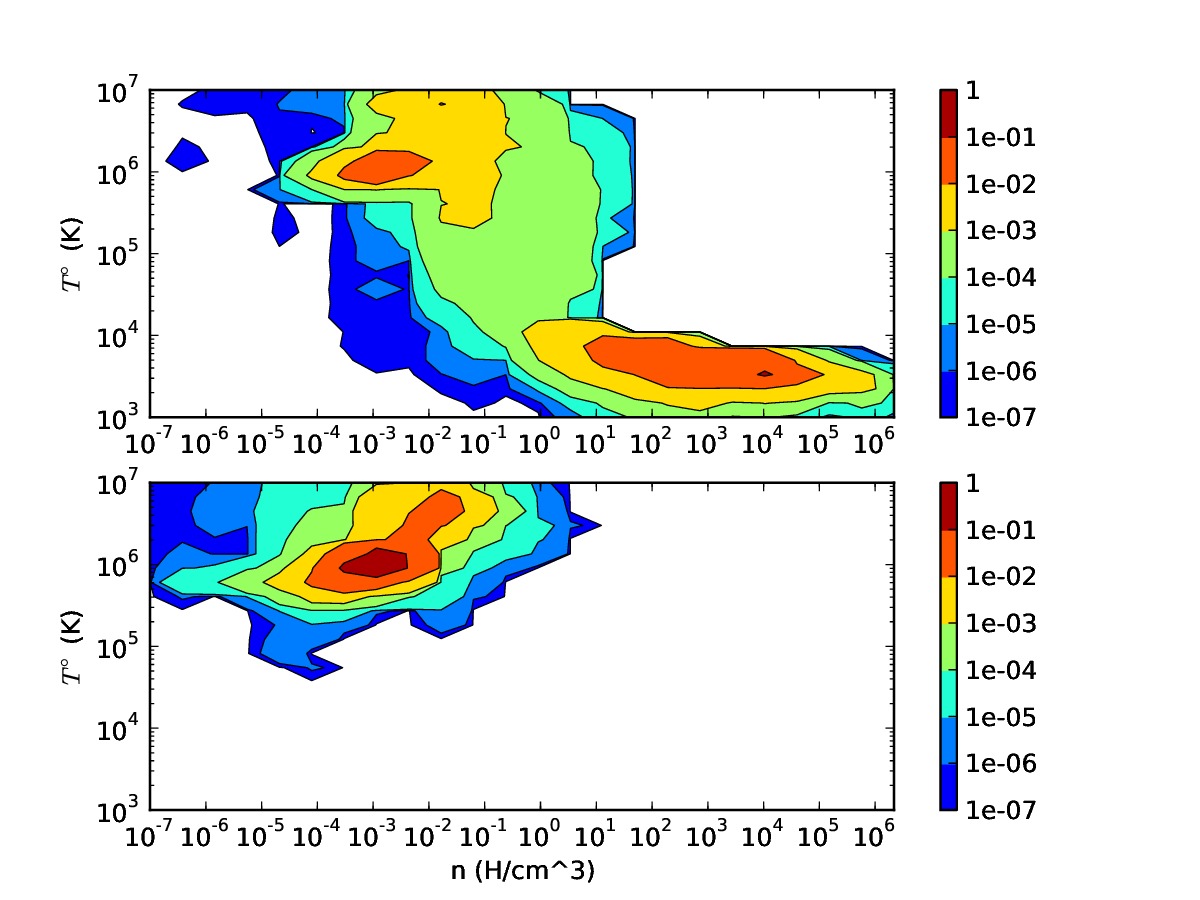}
  \caption{As in Fig. \protect{\ref{fig:density_temp1}} but for the
    simulation including hot coronal gas
    initially. \label{fig:density_temp2}}
\end{figure*}

Figure \ref{fig:face_on} shows a sequence of density cuts through the
disk plane.  The density scale on the left of each panel describes the
colour code for density in the range $n \in [1.5\times10^{-5},
1.5\times10^{4}]$ atoms/cc. The hot gas pushing through the cold disk
from below picks up a small amount of angular momentum.  This, in
turn, creates a strong shear layer in the disk plane.  This shear
gives rise to a rampant Kelvin-Helmholtz (KH) instability.  The KH
waves develop and break, giving the familiar saw tooth pattern.  The
toothed ring then fragments: the inner parcels continue to move inward
and the outer parcels move outward consistent with their specific
angular momentum values.  Gas from the destroyed ring coalesces into
new rings and the process repeats.  This limit-cycle behaviour is seen
as saw-tooth-like wiggles in the angular momentum contours in Figure
\ref{fig:gas_angmom}.  At approximately 200 Myr into the wind-impact
event, a stellar bar forms.  The influence of the bar on the flow
pattern is noticeable at $T\approx 300$ Myr.

The extent of the multiphase medium is shown in Figure
\ref{fig:density_temp1} for the disk plane (tow row) and the halo
(bottom row).  Recall from Paper 1 (Section 3.2.1) that the simulation
code using a multiple time step algorithm.  The left-hand column shows
the gas after one major time step $\delta T$ (25 Myr), chosen to allow
the initial conditions to come to an approximate local equilibrium.
The right-hand column shows the state of the gas after 200 Myr.  These
density--temperature plots show the classically expected two-phase
instability \citep{Field.etal:1969}.  The hot gas compresses the
initially cold in the inner disk and shortened cooling time keeps it
cold.  The initially cold gas in the outer disk is warmed by the
turbulent interaction with hot wind and this gas create the outer
sheath.  This results in the continuous distribution between the two
phases in the figure.  Note that this transitional gas only exists in
the vicinity of the disk (top row); the sheath itself has $T\approx
10^6$ K (bottom row).  Prior to the wind interaction, the $T$--$\rho$
phase plane shows the hot tenuous wind and cold disk gas in two
distinct and separated modes.

\begin{figure*}
  \centering
  \includegraphics[width=\textwidth]{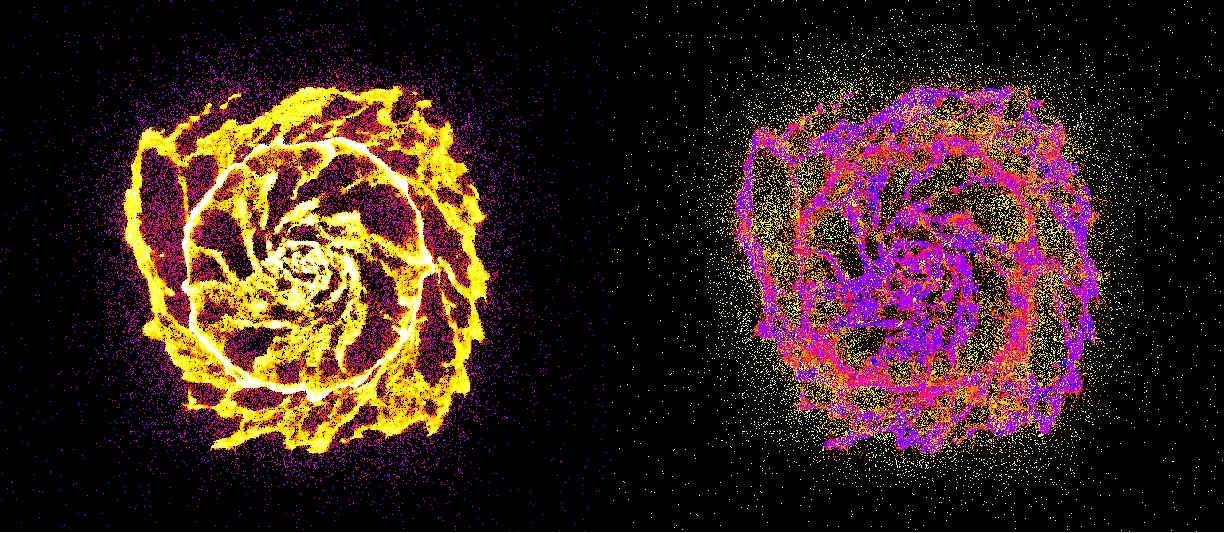}
  \caption{Cold disk from the simulation from
    Fig. \protect{\ref{fig:face_on}} at 100 Myr.  The left-hand panel
    shows gas density and the right-hand panel shows temperature.  In
    both panels, the scalar is logarithmically mapped to a colour
    sequence that transitions from blue to red to yellow to white.  On
    the left, the density of the hot gas (red) is $10^{-2}\cm^{-3}$
    and the density of the cold gas is $10^3\cm^{-3}$.  On the right,
    the host gas (yellow-white) has $T\approx10^6$ K and the cold gas
    (blue) has $T\approx5000$ K.  }
  \label{fig:face_on_images/dens_temp}
\end{figure*}

\begin{figure*}
  \centering
  \includegraphics[width=\textwidth]{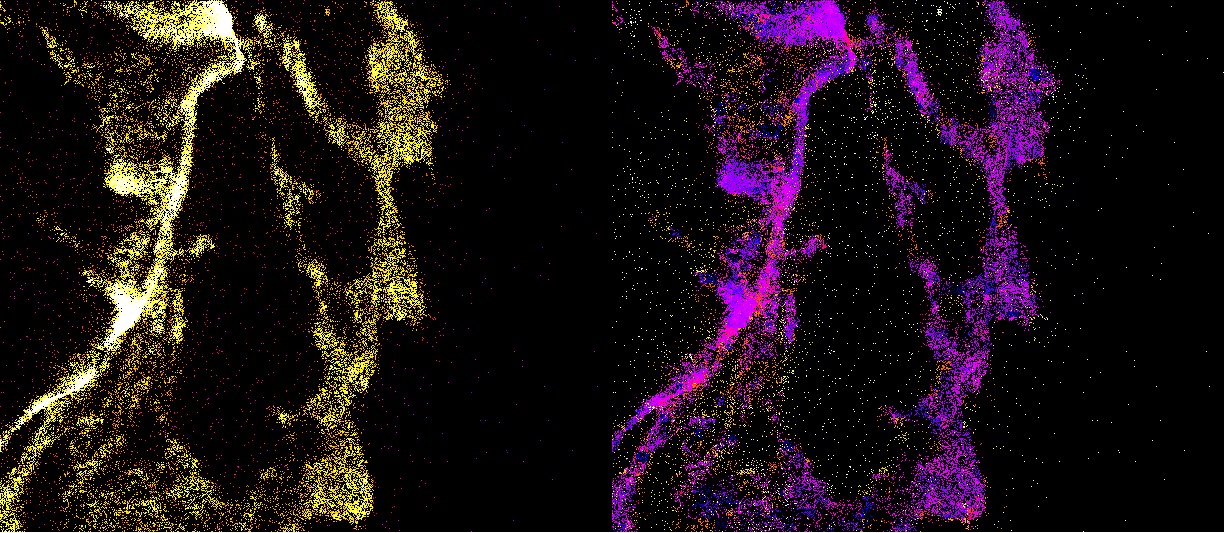}
  \caption{Cold disk at 100 Myr: close up.  Panels and scales as in
    Fig. \protect{\ref{fig:face_on_images/dens_temp}}.}
  \label{fig:face_on_close_up}
\end{figure*}

Figures \ref{fig:face_on_images/dens_temp} and
\ref{fig:face_on_close_up} show the density and temperature for the
gas at 100 Myr from the sequence in Figure \ref{fig:face_on}.  It is a
particular moment in the non-linear evolution of the KH instability
excited by shear created by the low angular momentum gas forcing its
way through the disk.  This particular snapshot shows a combination of
coalesced rings connected by the spokes of the remnant KH instability.
A detailed check shows that the hot and cold phases are in approximate
pressure equilibrium.  Moreover, the classic saturated KH
instabilities are clearly seen on multiple scales (i.e. examine the
saw-tooth shaped interface between the high and low density gas in
these figures).  Unlike previous figures that render ensemble-averaged
field quantities, this figure shows individual gas simulation
particles colour coded by their ensemble field values.

\begin{figure*}
  \centering
  \includegraphics[width=\textwidth]{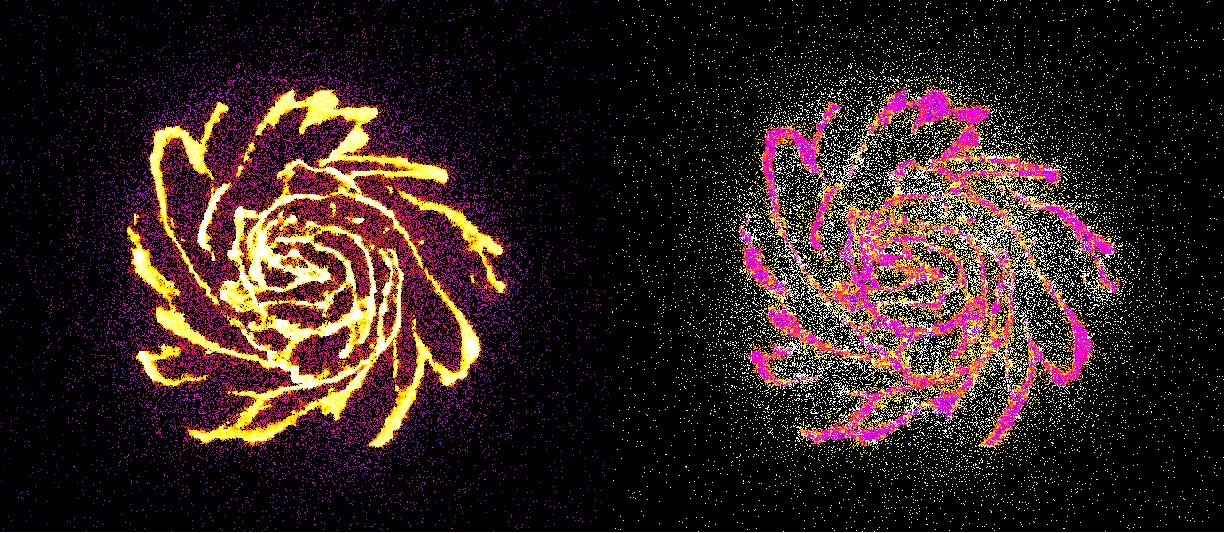}
  \caption{Cold disk at 200 Myr.  Panels and scales as in
    Fig. \protect{\ref{fig:face_on_images/dens_temp}}.  Note that the
    hot gas sits in the \emph{void} opened up by shear in the cold
    gas.  The saturation of the KH instability gives rise to rings and
    spokes, similar to those seen in ring galaxies.  Moreover a close
    examination of the ring in the Cartwheel reveals the same sort of
    scalloping seen in the simulation shown here. }
  \label{fig:200_close_up}
\end{figure*}

Finally, we may ask, are such things seen in Nature?  I would argue
that they are.  For example, compare NASA/ESO image of the Cartwheel
galaxy\footnote{e.g. \url{http://chandra.harvard.edu/photo/2006/cartwheel/}}
with the simulation profiles at 200 Myr (Fig. \ref{fig:200_close_up}).
Commonly, these rings and spokes are interpreted as a \emph{splash}
created by an intruder.  A close look Cartwheel image galaxy reveals
the same spokes and KH \emph{scalloping} seen in Figure
\ref{fig:200_close_up} suggesting that the shear between hot wind gas
and the cold disk gas is shaping the in-plane features!  ROSAT and
Chandra observations of the Cartwheel system reveal a stream of
soft-x-ray emitting gas connecting the Cartwheel to its apparent
companion intruders \citep{Wolter.Trinchieri:2003} with an inferred
column density of $N_H=1.3\times10^{21}\cm^{-2}$.  Assuming an outer
ring diameter of 60 kpc, one estimates a hot gas space density of
roughly $n_H=5\times10^{-3}\cm^{-3}$, comparable to the wind densities
in the simulations here.

\citet{Higdon:1996} estimated the expansion rate of the ring to be
$53\pm9\kms$ based neutral H observations, consistent with the passage
of an intruder approximately 300 \myr\ ago.  \citet{Amram.etal:1998}
estimated a mean expansion velocity of $13$--$30\pm10\kms$ with a
superimposed sinusoidal distortion of amplitude $20\pm5\kms$ from
H$\alpha$ observations.  However, the expansion rate of the ring is
similar to that found in the non-linear KH instability due to the
wind.  Altogether, it seems plausible that the observed spokes are
not, as some have suggested, simply the original spiral arms of the
target galaxy reappearing but the same KH instability seen in the
simulations described here.

\subsection{An edge-on wind}
\label{sec:edgeon}

The affect of the edge-on wind differs qualitatively and
quantitatively from that of the face-on wind.  First, as might be
expected, the shock forms in a rim around the outer cold disk, not
like a Mach disk.  The post-shock gas mixes with and ablates the cold
gas. The angular momentum of the initially cold gas gives the flow
some spin relative to the disk. The mixture of incoming wind and
ablated cold gas moves around the disk in the direction of disk
rotation. Approximately 70\% of the cold disk's initial angular
momentum content is lost to ablation in this way.  However, only
approximately 10\% of the cold disk mass is lost by 500 Myr after the
encounter; this is nearly all of the gas outside of a radius 0.04 (12
Kpc in Milky Way units).  Unlike the resulting disk in face-on
encounter, there is little shrinkage of the inner disk.

At the same time, the hot wind flows under and over the disk and
subsequently this gas is gravitationally focused on to the face of the
disk.  This, in turn, pressurises the cold disk and drives a strong
multiphase instability. The pressure from the gravitationally focused
wind is higher than from the Mach-disk-protected face-on wind leading
to higher density in the cold gas phase.  The phase-plane distribution
is similar to that for the face-on wind but very little of the gas has
transitional densities and temperatures; the two phase remain well
separated.  In summary, although the overall outcome is similar to
that of the face-on interaction, many features of the dynamical
pathway to the final state are different.

\begin{figure*}
  \includegraphics[width=0.9\textwidth]{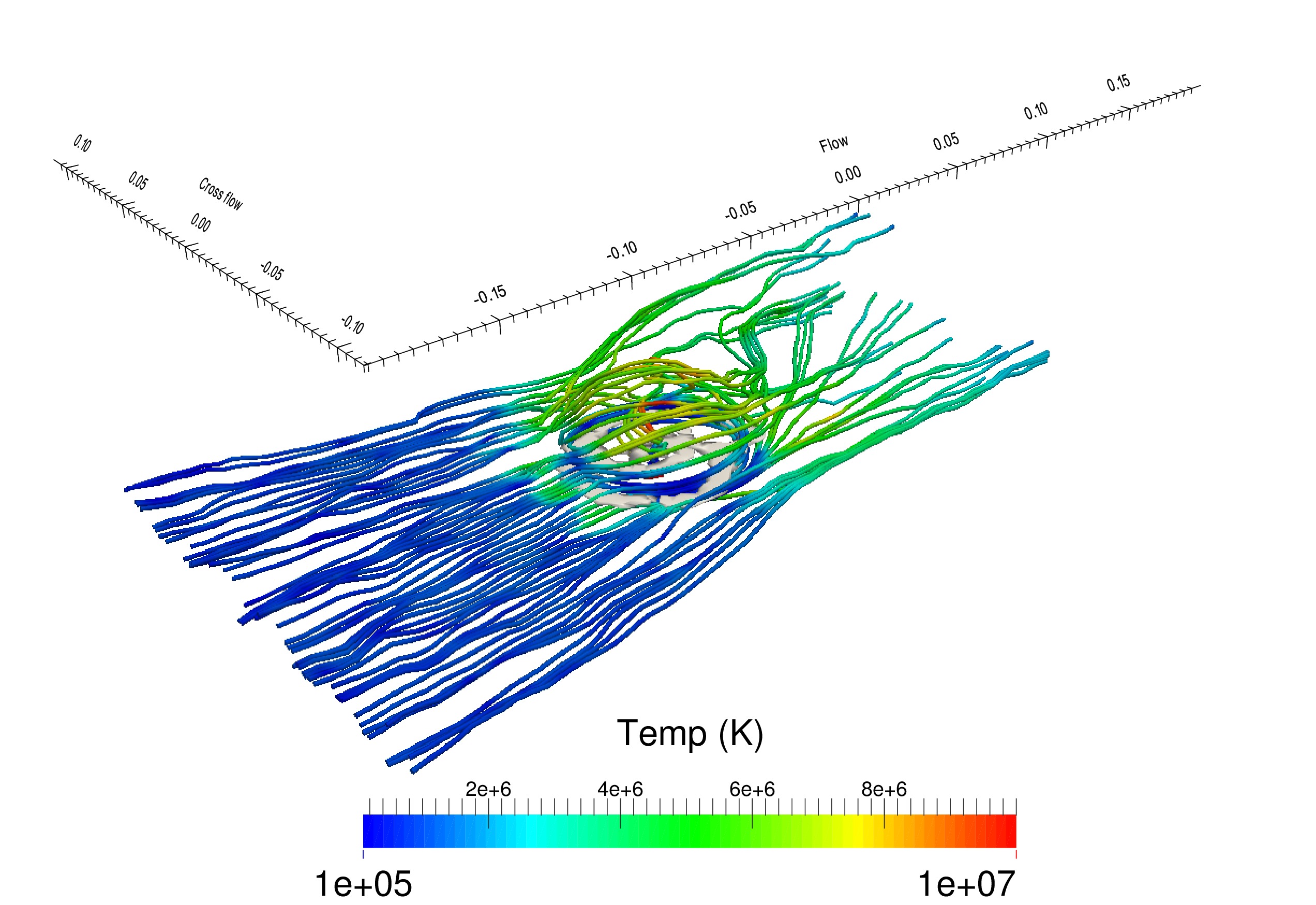}
  \caption{Streamlines showing gas flow for an edge-on impact of a hot
    wind with a cold gas disk at 400 Myr after impact.  The wind flows
    from left to right in this figure and is colour coded by gas
    temperature (K).  The gas density of the cold disk layer with
    density of 1 atom/cc is shown as the white surface.  The up-stream
    gas is gravitationally focused toward the galaxy. The gas shocks
    on impact and the hot flow ablates the cold gas as in the face-on
    impact.  Here, the ablated gas is pushed over or under the disk;
    through mixing, the mixture of the wind gas with disk gas gives
    the flow some overall rotation. Some of the hot gas is refocused
    down to the disk plane, heating the disk from above.}
    \label{fig:edge_on_stream}
  \end{figure*}

\begin{figure*}
  \includegraphics[width=0.8\textwidth]{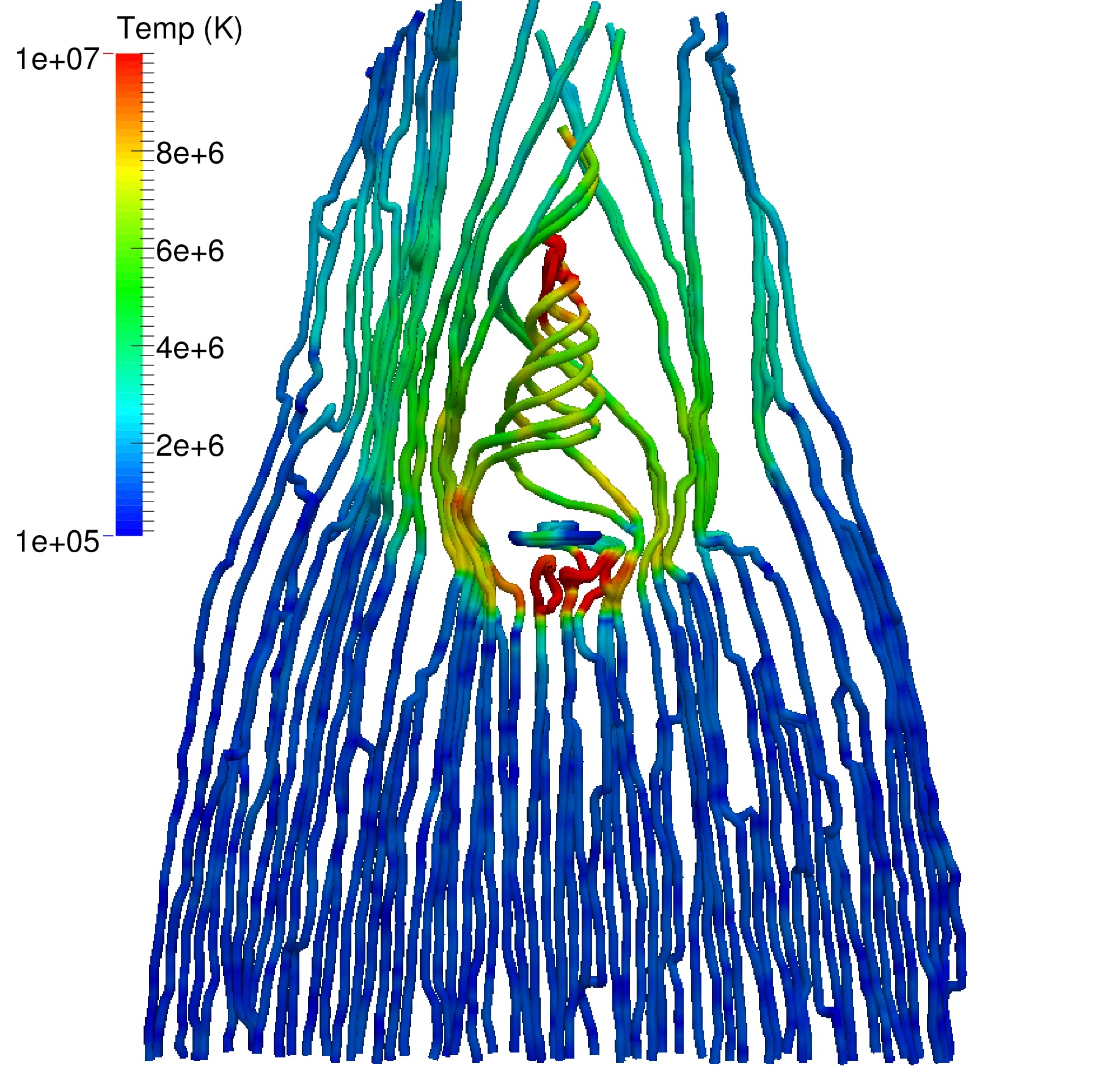}
  \caption{Similar to Fig. \ref{fig:gas_struct}, streamlines showing
    gas flow for an edge-on impact of a hot wind with a cold gas disk
    at 250 Myr after impact.  The flow is qualitatively similar to
    that without a corona.  The main difference is the higher
    temperature of the downstream flow at small cylindrical radius.}
  \label{fig:stream_corona}
\end{figure*}

\subsection{The addition of coronal halo gas}
\label{sec:hothalo}

To get some insight for the overall change in the wind--cold-disk
interaction caused by a hot coronal halo, we repeat the simulation
described in detail in Section \ref{sec:gal_hard} but adding a hot
isothermal coronal component in hydrostatic equilibrium in the galaxy
with an initial temperature of $10^6$ K and a density of $10^{-4}$
atom/cc at a radius of 30 kpc.  This value was chosen to match typical
hot halo densities estimated at large galactocentric radii.  However,
as described by \citet{Fang.etal:2013}, this value may overestimate
the inner galaxy density and underestimate the total coronal gas mass
owing to the assumption of a cuspy, NFW profile.

The main results are as follows. The pressure from the hot-wind impact
is more readily communicated through the coronal halo that is already
in place downstream of the shock.  This increases the pressure on the
inner disk and results in a higher amplitude multiphase instability.
This is seen through comparison of the density-temperature phase plane
at the same interaction time with and without a corona (compare
Figs. \ref{fig:density_temp1} and \ref{fig:density_temp2}).  Recall
that the top row shows the gas in the disk and the bottom row shows
the gas in the halo.  The initial state in the vicinity of the disk
reveals the superposition of the hot, hydrostatically-supported
higher-density ($\gtrsim3\mbox{H}/\mbox{cm}^3$) component peaking in
the inner disk at and a lower-density
($\gtrsim3\times10^{-2}\mbox{H}/\mbox{cm}^3$) component in the outer
disk.  Overall one sees more high density cold gas in the simulation
with a corona.  In addition, one sees that there is a smoother
transition between the cold-dense and hot-tenuous phases in the flow
without a corona; the corona suppresses the production of the
intermediate-state gas by providing an additional heat source.

\begin{figure}
  \begin{minipage}{0.475\textwidth}
    \includegraphics[width=\textwidth]{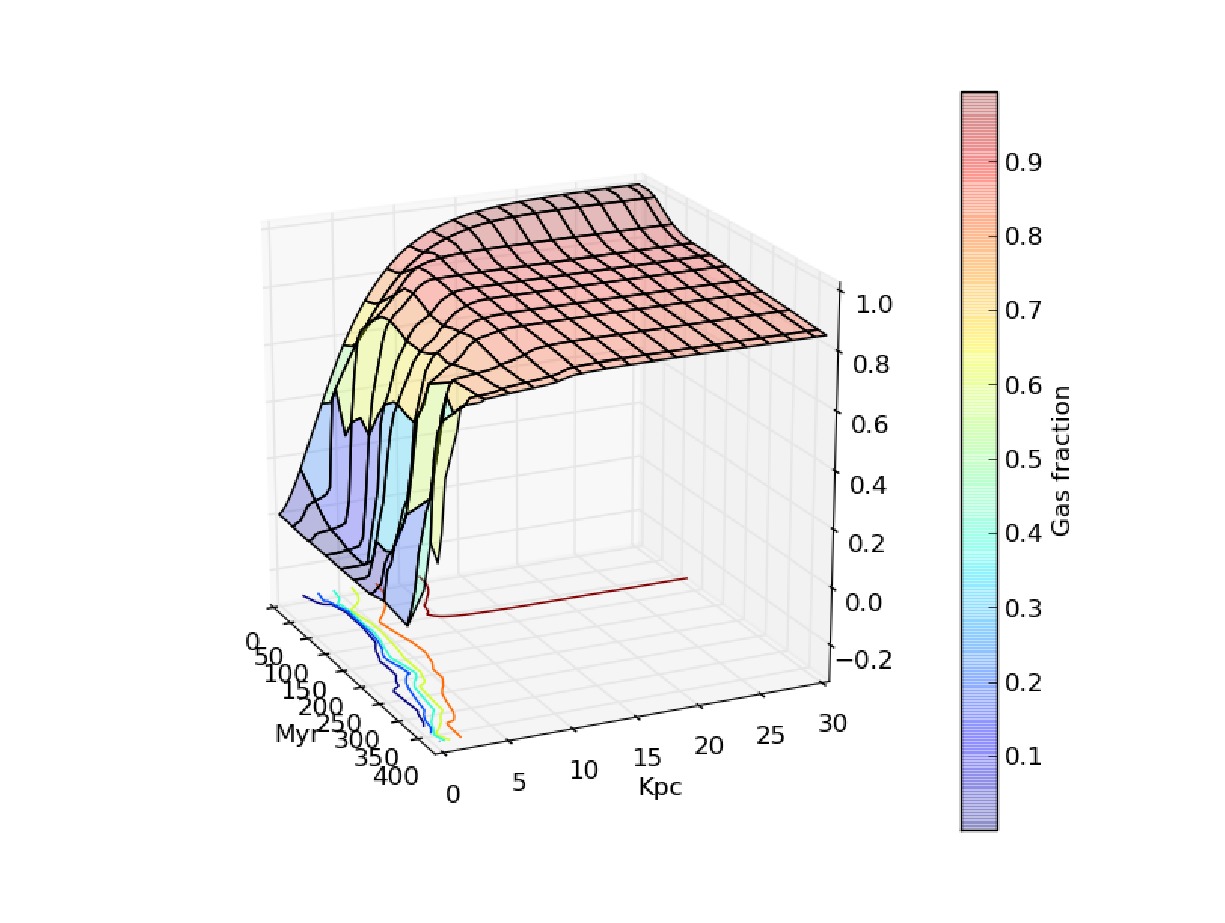}
    \caption{The cumulative fraction of gas in the disk with radius (in
      kpc) and time (in millions of years) for simulation with a
      coronal halo (cf. Fig. \protect{\ref{fig:gas_fraction}}). }
    \label{fig:gas_fraction2}
  \end{minipage}
  \hspace{0.05\textwidth}
  \begin{minipage}{0.475\textwidth}
    \includegraphics[width=\textwidth]{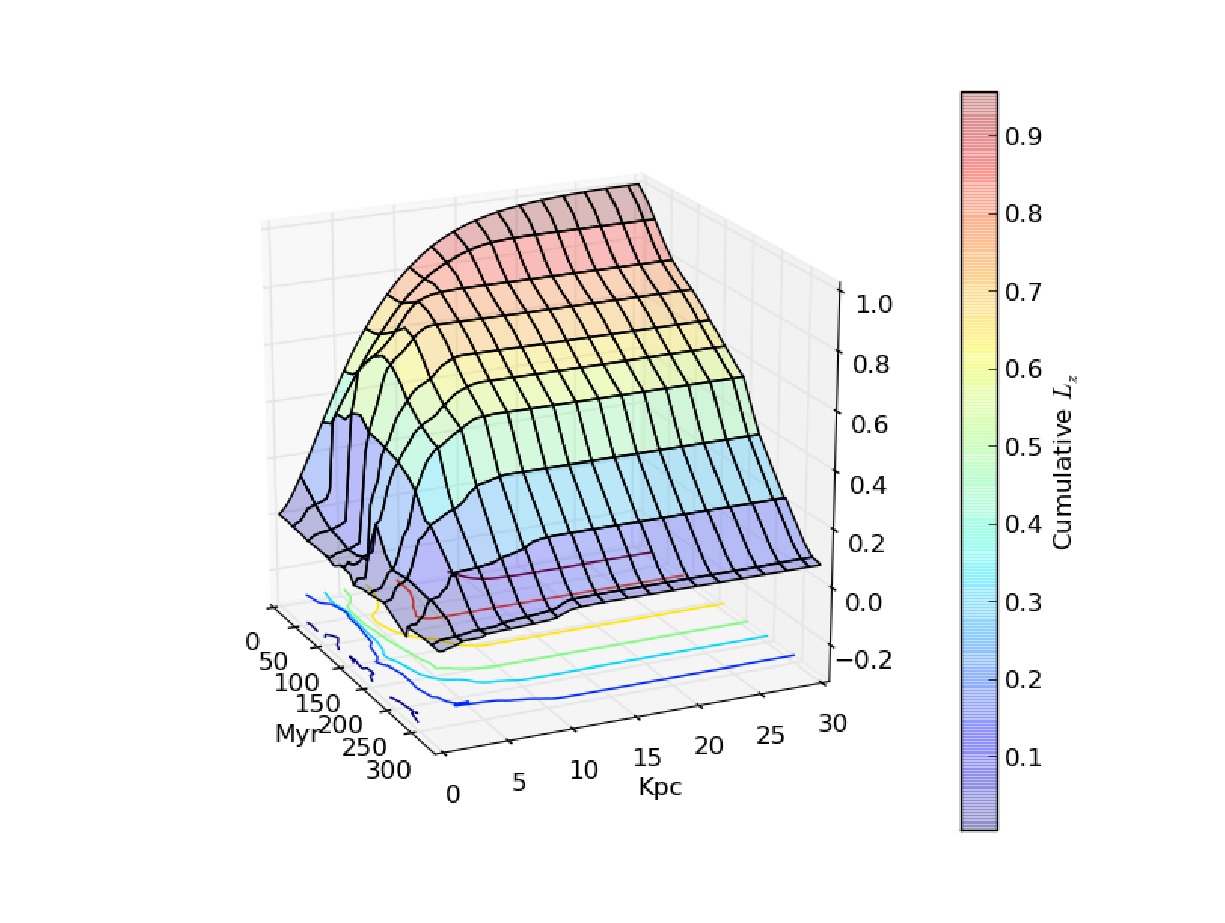}
    \caption{Evolution of the cumulative angular momentum
      fraction of gas in the disk with time for simulation with a
      coronal halo (cf. Fig. \protect{\ref{fig:gas_angmom}}).}
    \label{fig:gas_angmom2}
  \end{minipage}
\end{figure}

Similarly, the higher density of the hot gas at smaller galactocentric
radii increases the angular momentum transfer from the cold disk to
the post-interaction outflow by increasing the collisions of the hot,
low angular momentum gas and the cool, high angular momentum gas.
This is seen in the comparison of the cumulative angular distributions
shown in Figures \ref{fig:gas_angmom} and \ref{fig:gas_angmom2}.  The
total angular momentum drops nearly linearly to 10\% of its initial
value at 450 Myr after the initial impact.  However, at the same time,
the removal of this angular momentum causes the cold gas disk to
shrink.  This decreases the overall mass loss.  In other words, the
net result of the coronal gas is to \emph{protect} the initially cold
gas mass against stripping through ablation (compare
Figs. \ref{fig:gas_fraction} and \ref{fig:gas_fraction2}). The disk
loses only approximately 20\% of its total mass overall, but by 450
Myr, all of this gas is inside of 8 kpc.

The overall flow pattern is the same with the hot corona
(Fig. \ref{fig:stream_corona}) and without (Fig. \ref{fig:gas_struct},
lower panel).  The extent of both the inner and outer density sheaths
are smaller with the corona, consistent with larger angular momentum
loss and disk shrinkage.  Moreover, the simulation clearly shows that
the density of the cool gas in the inner disk is much higher than
without the coronal gas.  The effect of the coronal gas on the inner
galaxy may itself place a limit on the density of the inner coronal
gas and thereby also place a limit on the shape of the galactic dark
matter potential.  This scenario is ripe for further investigation.
Use of the non-LTE DSMC code, under development, will allow prediction
of line-strengths for observed density and temperature diagnostic
features (e.g. Mg II absorption).

\section{Discussion}
\label{sec:discuss}

The effect of ram pressure on galaxy morphology and transformation has
been studied in several contexts over the last decade.  A particularly
nice example is provided by \citet{Crowl.etal:2005} who inferred the
signature of ram pressure and ablation by combining optical, H{\sc I},
and radio continuum observations for NGC 4402 (cf. Figs. 2 and 3 from
\citealt{Crowl.etal:2005}).  They argue that the absence of reddening
at the disk's edge implies truncation and that the position angle of
linear dust filaments coincides with radio continuum, suggesting the
ICM wind orientation.  Similarly, \citet{Vollmer.etal:2008} show that
observed regions of extraplanar gas with H-$\alpha$ flux have larger
velocity than those with CO flux in Virgo galaxy NGC 4438, suggesting
that the ionised gas is excited by the ram-pressure mechanism.  All in
all, there is little doubt that ICM winds influence a galaxy's ISM.

These and related observations have motivated a variety of
computational hydrodynamic efforts and provide opportunities for
comparison. The morphology of the flows in our simulations have the
same morphological features as those performed with hydrodynamic codes
\citep[e.g.][see his Figs. 2 and 7, using GRAPE-SPH]{Bekki:2009}.  We
expect DSMC to have lower resolution for the particle numbers used
here than many of the hydrodynamic simulations discussed above, and
therefore the good correspondence is encouraging.  Also as expected,
the interfaces between gas components and multiple phases are
distinct, demonstrating a virtue of the kinetic approach.

However, since most authors had different sets of interests, goals,
and physical choices, direct intercomparisons are often approximate.
In addition to varied initial conditions, many used different
equation of state combinations with and without atomic and plasma
cooling.  For some examples, \citet{Quilis.etal:2000} simulated the
interaction between a spiral galaxy and the ICM of a rich cluster at a
relative speed of 2000 km/s.  They show that the interaction removes
most of the gas in 100 Myr and argue that cooling is not important in
this case.  Our wind speeds are nearly an order of magnitude smaller
than this, yet encouragingly, the morphological features of the
ram-pressure response are very similar to those found here (see their
Fig. 1).  In contradiction, we conclude that the wind--disk
interaction cannot remove all of the initially cold gas.  However, we
have not tried such a fast encounter; given the greatly increased
energy and momentum content in this fast wind, it seems plausible that
our result would be similar to theirs.  \citet{Marcolini.etal:2003}
and \citet{Mayer.etal:2006} performed hydrodynamic simulations of ram
pressure stripping for dwarf galaxies.  The former investigation
considers winds typical of poor galaxy groups and considers the role
of mass, velocity and wind direction.  They find that dwarf galaxies
may be stripped in a few hundreds of $\myr$ (see their Fig. 4 for an
illustration of the interaction morphology).  \citet{Mayer.etal:2006}
pattern their study on a Milky-Way-like system and consider the
combined roles of ram pressure and orbitally induced tidal forces.
\citet{Boselli.etal:2009} attempted galaxy evolution modelling with ram
pressure and starvation.  Their work suggests that anaemic spirals with
truncated gas distributions and young populations have been produced
by recent ram pressure event and cannot be produced by starvation
alone.  Similarly, they report that low luminosity dwarf disk galaxies
can be transformed in dE galaxies with ease.  Most of these are not
directly comparable to the simulations described in
Section \ref{sec:gal_hard} but they still exhibit many of the same
morphological features qualitatively.

Many researchers compare their stripping results to the predictions of
the Gunn--Gott criterion, although the general consensus is that this
is only a crude approximation.  In particular, as in this paper, a
number of hydrodynamic simulations indicate an early phase of rapid
stripping in the outer disk followed by a period of slower, viscous,
stripping \citep[e.g.][]{Marcolini.etal:2003, Roediger.Hensler:2005}.
However, we demonstrate here that the initial mode of rapid stripping
is ablation, not impulsive momentum transfer.  Based on an extensive
parameter survey, \citet{Roediger.Hensler:2005} develop a revised
criterion based on pressure at the disk mid plane.  However, it is
clear that the mass loss in the end depends on the widely differing
ICM densities, velocities and galaxy mass \citep{Bekki:2009}.

In this study, our goal was the interaction physics itself rather than
the implications for a specific astronomical scenario.  In summary,
although momentum is transferred from the wind to the disk ISM, the
process bears little resemblance to its moniker, ram pressure, but
rather seems to be an elaborate interplay between shocks, shear
fronts, kinetic ablation (i.e. evaporation) and angular momentum
transfer.  Although it is natural to seek a scaling relation such as
the Gutt-Gott criterion to summarise the overall outcome, the results
of this paper suggest that an impulsive ram pressure approximation
will be crude at best.  The large variation of successes and failures
of the criterion reported in the literature is not a surprise in this
light.

Motivated by he discovery of a large ($M\approx3\times10^8\msun$) HI
cloud in the Virgo cluster that appears to originate from the galaxy
NGC 4388 \citep{Osterloo.vanGorkom:2005}, \citet{Roediger.etal:2006}
used hydrodynamic simulations to confirm that ram pressure is
sufficient to strip this galaxy's gas mass.  They report a distinctive
flow pattern that they ascribe to von Karman oscillation, typically
produced by the unsteady flow of a fluid past a blunt body, and
suggest that the \citet{Osterloo.vanGorkom:2005} discovery is evidence
of this.  The authors do remark that the observed HI tail morphology
does not match the simulation and suggest that this discrepancy is due
to missing physics and ICM interaction.  We agree that this is likely.
In particular, the results of this paper challenge the assumption that
an adiabatic gas \citep{Roediger.etal:2006} will correctly capture the
nature of the mesoscopic interactions important at interfaces and that
the hydrogen lost from the stripped galaxy will be neutral.  As in the
present work, \citet{Roediger.Bruggen:2006} found that wind
orientation angle does strongly affect mass loss unless the wind
direction is nearly edge on.

\citet{Tonnesen.etal:2007} studied the problem from the $\Lambda$CDM
galaxy formation and evolution viewpoint by tracking galaxies in
clusters over time in a cosmological volume to ascertain the
distribution of evolutionary drivers. They conclude that gas stripping
is the most common mass loss mechanism and remains important out to
the virial radius of the cluster. This general finding is consistent
with our simulations.  However, their aim was an ensemble survey using
$\Lambda$CDM initial conditions.  Given their limitations on
resolution, a direct comparison with the details of the dominant
physical mechanisms in their simulations is not possible.

Finally, \citet{Kapferer.etal:2009} explicitly investigated the
dependence of star formation and morphology on the strength of ram
pressure using combined N-body and hydrodynamic simulations with
recipes for star formation and stellar feedback.  They find that star
formation is enhanced by more than a magnitude in the simulation with
a high ram pressure relative to the same system evolving in isolation.
They report that newly formed stars in the disturbed gas stars fall
back to the old stellar disc, building up the bulge. They find also
that gas stripping saturates quickly as in the simulations reported
here.  In addition, they describe a complex velocity pattern due to
the rotation and spiral arms. This is qualitatively identical to the
KH instabilities described in Section \ref{sec:gal_hard}.

All in all, the simulations reported in this paper describe similar
features and mass loss fractions to those from hydrodynamic
simulations when direct comparisons are possible.  Moreover, our
simulations underscore the importance of the interface in predicting
the evolutionary morphology of ram-pressure induced flows and
response.  This suggests that the kinetic approach may become an
essential tool for clarifying the dynamics of multiscale gas
interactions for astrophysical flows.

\section{Summary}
\label{sec:summary}

The ram pressure mechanism creates flows in multiply distinct gas
phases with dynamically important interfaces, such as shock fronts and
phase boundaries.  Resolving the physics at these interfaces requires
subtlety when using a hydrodynamic method but is naturally done by
direct solution of the collisional Boltzmann equation using a
kinetic-theory method.  To this end, we explores the use of the hybrid
kinetic theory--n-body code from Paper 1 to provide a check on our
physical understanding of the ram-pressure scenario.  The kinetic
theory algorithm, Direct Simulation Monte Carlo (DSMC), is
computationally costly, so we have not extensively surveyed the range
of parameters relevant to all ICM--ISM interactions but rather focused
on the dynamical mechanism for a single case, more typical of a group
ICM density and velocity for a Milky-Way like galaxy
\citep{Freeland.Wilcots:2011}.  Overall, the findings from the
simulations here are consistent with many of the previously published
results.  The simulations include a gravitationally self-consistent
stellar disk, spheroid, and dark matter halo with an initially cold
gas layer.  The wind temperature is $10^6$ K with a uniform bulk
velocity set to twice the virial velocity of the galaxy initially.
The choice of a group rather than rich cluster environment is partly
motivated by computational expense.  The key physical features of the
gas response are as follows:
\begin{enumerate}
\item The incoming wind shocks as it approaches the disk and a
  compression wave begins to propagate upstream. Meanwhile, the
  post-shock gas is heated, compressed, and this results in a lower
  bulk velocity in the downstream gas by approximately a factor of
  four.  The ballistic crossing time of the hot gas is now close to
  the disk crossing time.  The velocity of the hot wind gas decreases
  further as it pushes its way through the multiphase ISM.
\item The cold gas is not launched from disk en masse as in the
  impulsive approximation \citep{Gunn.Gott:1972}, but rather, the hot
  post-shock gas ablates and entrains the initially cool disk gas.
  The stripped gas ends up with a temperature close to that of the
  entraining wind.
\item The hot wind, either pushing through the disk for a face-on and
  oblique flow or impacting the disk from below and above for an
  edge-on flow, pressurises the cold disk and promotes a two-phase
  instability.  The rapid consumption of gas by star formation and
  removal of the cold gas supply may well result in morphological
  transformations as previous authors have suggested. Indeed,
  \citet{Kapferer.etal:2009} and \citet{Tonnesen.Bryan:2010} find that
  gravothermal instability may promote cooling and star formation; our
  simulations neither include the molecular cooling or gravity on
  molecular cloud scales needed to investigate this possibility.
\item In the face-on or oblique wind--disk encounters, the post-shock
  ICM flowing through the now porous ISM has low specific angular
  momentum relative to the rotationally supported ISM.  This results
  in a strong shear between the two phases with two important
  consequences: 1) the shear excites a strong Kelvin-Helmholtz
  instability that results in ``ring and spoke'' morphology; and 2)
  the hot wind carries off some the gas disk's angular momentum.  This
  results in an inward advection of cold gas with the possible
  formation of an inner ring and a net rotation of the wind escaping
  the disk.  The addition of angular momentum to the coronal plasma
  with the rotational orientation of disk may have interesting
  implications for galactic dynamos.
\item Since the escaping gas is primarily lost from the outer disk,
  the outflow appears as a bell-shaped sheath for the face-on and
  oblique flows.  The angular momentum gained through the interaction
  appears to help collimate the sheath by providing a centrifugal
  barrier at small radii.  For an edge-on encounter, there is no
  well-defined inner sheath, although the cold gas disk casts a
  geometric shadow on the down stream flow creating a density dip
  along the flow axis.
\end{enumerate}

In summary, we find that some initially cold gas does leave the galaxy
as do most studies, but our simulations suggest that the process is
complex. It is, therefore, unlikely be shoehorned into a simple
scaling law owing to the subtle dependence on initial conditions.  The
work here further suggests that the gas leaving the galaxy due to the
wind interaction will be \emph{hot} (i.e. not detectable as neutral
hydrogen).  We would need higher resolution and huge particle numbers
to see the possible entrainment of small cold clumps or to resolve
gravothermally unstable clumps.  Furthermore, these simulations show
that radiative cooling is critical to the flow pattern and cannot be
ignored. Without cooling, most of the cold gas would heat and
subsequently evaporate whereas with the cooling, we find strong
multiphase instabilities that protect the gas against further
large-scale mass loss.  More generally, the phase and shock interfaces
are likely to control the overall evolution of the gas and produce
structure at all scales!  Similar to collisionless dynamics, one can
not gloss over interface regions even if their total volume is small.

Although DSMC was developed for high-Knudsen number (i.e. rarefied)
flows, it may be used for near continuum flows as well.  DSMC is more
computationally expensive than traditional hydrodynamics for
near-continuum flows, but has the advantage of unconditional
stability, self-consistent momentum transport and no difficulties with
formal discontinuities\footnote{In this sense, one might say that DSMC
  is 100\% shock capturing, although more precisely DSMC has no need
  to \emph{capture} a discontinuity because the physics of a shock
  front is resolved by the algorithm at the scale of the effective
  mean free path.}.  In particular, an ideally tuned DSMC simulation
simultaneously maintains approximately 10 particles per simulation
volume whose length scale is approximately 1 mean-free path.  For
denser flows, this condition is computationally infeasible.  The
implementation described in Paper 1 has a collisional limiter that
transitions to the near equilibrium solution of the energy and
momentum flux equations when the collision rate becomes very high.
However, even in this limit, momentum and energy remain precisely
conserved and numerics of the flow remains stable.  This makes DSMC
ideally suited for investigating flows with momentum transfer and
heating or cooling by conduction or radiation at interfaces.

Since this project arose as a code test, we have chosen to use an LTE
approximation for gas cooling; this approach is also known as the
\emph{quasi static state} (QSS) approximation.  QSS describes that
collisional and radiative excitations as rates based on the local
field quantities.  In addition, we do not include electrons as a
separate species for these simulations, but include their influence in
the QSS limit.  Therefore, electron conduction is not
self-consistently simulated although hard-sphere conduction of the
atoms is.  On the other hand, a precise study of the excitations and
electron interactions at the interfaces is necessary for estimating
their physical state and dynamical influence on the flow.  Moreover,
accurate observational predictions depend on the state of the gas at
the interface.  It is precisely such investigation of interface
dynamics and energetics on small and intermediate scales that
motivated the combined DSMC--n-body code for transitional flows in the
first place.

Therefore to improve the self consistency of these simulations, we are
actively developing a version of our hybrid code that includes the
full CHIANTI\footnote{CHIANTI is a collaborative project involving
  George Mason University, the University of Michigan (USA) and the
  University of Cambridge (UK).}  atomic database
\citep{Dere.etal:1997,Landi.etal:2011} for collisional cross sections
and recombination (bound-bound and bound-free) and standard plasma
transitions (free-free) for the ionised regime. This new
implementation will allow us to straightforwardly make explicit
predictions for both continuum and line-excitation strengths. However,
it does not yet include the electrons as a separate kinetic species,
but rather assumes that they follow the ions.  This restriction will
be relaxed in a future version that will allow a more realistic
treatment of electron conduction and the dynamical influence of fixed
magnetic fields.  To be sure, simulation of these processes based on
kinetic theory are very expensive but, in principle, will yield
detailed observational predictions and accurate internal energetics
and momentum transfer.

\section*{Acknowledgements}

This material is based upon work supported by the National Science
Foundation under Grant No. AST-0907951.  I thank Aeree Chung for
fascinating me with the perplexing dynamics of ram pressure physics.

\bibliographystyle{mn2e}

\label{lastpage}

\end{document}